\documentclass[12pt]{article}
\usepackage{amsmath,amssymb,epsfig}

\newcommand{\preprintnumber}[1]{\vbox{ \baselineskip 14pt \hfill
    \hbox{\normalsize } \\
\hfill \hbox{\normalsize #1} } \vskip 2cm}
\newcommand{\email}[1]{\footnote{email:#1}}

\newcommand{\C}{\mathbb{C}}
\renewcommand{\P}{\mathbb{P}}

\newcommand{\cO}{\mathcal{O}}

\renewcommand{\S}{{\rm S}}
\newcommand{\R}{{\mathbb R}}
\newcommand{\Z}{{\mathbb Z}}

\newcommand{\nn}{\nonumber}
\newcommand{\be}{\begin{equation}}
\newcommand{\ee}{\end{equation}}
\newcommand{\bea}{\begin{eqnarray}}
\newcommand{\eea}{\end{eqnarray}}

\usepackage{color}
\input{colordvi.tex}

\begin{document}

\title{\preprintnumber{KIAS-P12020}
$U(n)$ Spectral Covers from Decomposition}

\author{
Kang-Sin Choi$^{\rm a,b}$\email{kangsin@ewha.ac.kr} \ and Hirotaka Hayashi$^{\rm a}$\email{hayashi@kias.re.kr}\\
\it \normalsize $^{\rm a}$School of Physics, Korea Institute for Advanced Study, Seoul 130-722, Korea
 \\ \it \normalsize $^{\rm b}$Scranton College, Ewha Womans University, Seoul 120-750, Korea
}

\date{}

\maketitle

\begin{abstract}

We construct decomposed spectral covers for bundles on elliptically fibered Calabi--Yau threefolds whose structure groups are $S(U(1) \times U(4))$, $S(U(2) \times U(3))$ and $S(U(1) \times U(1) \times U(3))$ in heterotic string compactifications. The decomposition requires not only the tuning of the $SU(5)$ spectral covers but also the tuning of the complex structure moduli of the Calabi--Yau threefolds. This configuration is translated to geometric data on F-theory side. We find that the monodromy locus for two-cycles in K3 fibered Calabi--Yau fourfolds in a stable degeneration limit is globally factorized with squared factors under the decomposition conditions. 
This signals that the monodromy group is reduced and there is a $U(1)$ symmetry in a low energy effective field theory. To support that, we explicitly check the reduction of a monodromy group in an appreciable region of the moduli space for an $E_6$ gauge theory with $(1+2)$ decomposition. This may provide a systematic way for constructing F-theory models with $U(1)$ symmetries.

\end{abstract}

\newpage

\section{Introduction}

Compactification of string theory yields promising vacua related to Grand Unified Theories, based on various gauge groups, or directly to the Standard Model. String theory explains the origin of gauge theory with a large group such as $E_8 \times E_8$ from the consistency, but a background gauge field or vector bundle on the internal manifold can break that symmetry into a small and realistic one \cite{Candelas:1985en}. Therefore, constructing such vector bundle is a crucial step in understanding realistic vacua.

Spectral cover is one of the good tools for constructing an explicit vector bundle of heterotic string, automatically satisfying the supersymmetry condition for the vector bundle, 
if the compact manifold admits an elliptic fibration with a section that we call the zero section \cite{Friedman:1997yq, Donagi:1997dp, Friedman:1998si}. That construction has a direct application to a duality between heterotic string and F-theory \cite
{Friedman:1997yq, Bershadsky:1997zs}. Local components of the vector bundle are related to a set of points on the elliptic fiber,  spanning multiple covers over the base $S$ of the elliptic fibration. 
If the former has the structure group $SU(n)$ or $Sp(n)$, the corresponding points are simply described by a meromorphic function of the variables describing elliptic fiber. With an additional line bundle on the spectral cover, we obtain the desired vector bundle by the Fourier--Mukai transformation \cite{Friedman:1997yq}. 

In this work, we construct decomposed spectral covers for bundles involving non-unimodular structure group $U(1)$ or $U(n)$. Those decomposed spectral covers are important for three main reasons. First, many models have different matter contents with the same quantum number under the gauge group, which have to be distinguished. 
For example, in the $SU(5)$ Grand Unification, quantum numbers are the same for a quintet ${\bf \overline 5}$ containing Higgs doublet and one containing lepton doublets. They can be distinguished, if they are differently charged under another group, which may be gauged or global. This is the case if the group is further embedded to a larger unification group such as $SO(10)$ or $E_8$. In conventional heterotic and F-theory compactification, the components of a representation of the structure group $SU(5)_{\perp}$ are not distinguished or connected by a monodromy group, so we have to reduce the structure group \cite{Blumenhagen:2005ga, Tatar:2006dc, Donagi:2008kj,Bouchard:2009bu, Tatar:2009jk, Marsano:2009gv}. 
Realistic models may introduce an extra $U(1)$ such as the baryon minus lepton number or the Peccei--Quinn gauge group, which is embedded to larger unification group together with the gauge group and reproduce the desirable spectra. In other words, the structure group $SU(5)_{\perp}$ should be reduced to $S(U(1)\times U(4))$ or a subgroup thereof to yield such a $U(1)$ symmetry in a four-dimensional low energy effective theory \cite{realistic}. This class of groups will be extensively studied in this paper. Second, an unbroken $U(1)$ symmetry can also prevent dangerous operators in a low energy effective field theory, which can mediate unwanted nucleon decays. Moreover, we may want to directly construct models having not only simple but also with additional Abelian gauge group(s), to have $SU(3) \times SU(2) \times U(1)$ of the Standard Model \cite{SM} or $SU(5) \times U(1)$ of the flipped $SU(5)$ unification model \cite{FSU5}, for instance. They are respectively the commutants of $S(U(1)\times U(4))$ and $S(U(1) \times U(5))$ in $E_8$. In particular, since $U(1)$ is Abelian and commutes with itself, it may become gauge group if there is no St\"uckelberg mass term for the gauge boson from the Green--Schwarz mechanism \cite{Blumenhagen:2005ga, Tatar:2006dc}.

The construction of a decomposed spectral cover for an $S(U(1) \times U(m))$ bundle may guarantee the existence of a $U(1)$ gauge symmetry on the heterotic side, up to Green--Schwarz mechanism making it global\footnote{The $U(1)$ symmetry may be broken by the D-term conditions with a non-vanishing Fayet--Illiopoulos term \cite{Blumenhagen:2005ga, Tatar:2006dc}.}. 
However, there have been no real construction of such a spectral cover due to the following difficulties. First, on the elliptic curve, there cannot be a meromorphic function describing {\em only} one point as a zero. For this we just obtain a $U(1)$ spectral cover as a part of $SU(n) \; (m < n)$ spectral cover by tuning the defining equation. Second, locally, a spectral cover is multiple copies of the base $S$ up to a twist, but globally they may be connected by the monodromy from the structure group.
If we want to have an independent spectral cover which consists of one sheet and does not mix with other covers in the global sense, the location of the spectral cover should be not only globally well-defined and isolated but also different from the zero section. Therefore, not only the spectral cover is tuned but also the elliptic fiber $E$ in the heterotic side needs to be special, so that 
the position of the single cover correspond to a global section except for the zero section \cite{Hayashi:2010zp}. Likewise for an $S(U(n) \times U(n^{\prime}))$ spectral cover, we need a set of points whose sum under the group law on $E$, corresponding to the determinant $U(1) \subset U(n)$ or $U(1) \subset U(n^{\prime})$, should be again another global section which is not the zero section. This will be done in Section 2. We will explicitly construct decomposed spectral covers for bundles with $S(U(1) \times U(4))$, $S(U(2) \times U(3))$ and $S(U(1) \times U(1) \times U(3))$ structure group.

Even if we construct the $S(U(1) \times U(m))$ or $S(U(n) \times U(n^{\prime}))$ spectral cover on the heterotic side, the global existence of the $U(1)$ symmetry on the dual F-theory side seems not completely evident. 
The F-theory dual can be obtained by compactifications on K3 fibered Calabi--Yau fourfolds $X_4$ over the base $S$ in a stable degeneration limit, and the K3 itself has an elliptic fibration \cite{Vafa:1996xn}.  
In F-theory, a non-Abelian gauge group is realized by a singularity having the same 
Dynkin diagramatic structure after appropriate resolutions \cite{Morrison:1996na, Morrison:1996pp, Bershadsky:1996nh, Katz:2011qp}, classified by Kodaira. However, there is no singularity corresponding to a $U(1)$ symmetry in the classification, so we cannot observe nor construct it, for example, by the Tate's algorithm. 
A way to see it is by checking the existence of a certain type of topological $(1,1)$--forms of Calabi--Yau fourfolds, on which M-theory rank three antisymmetric tensor can give rise to the corresponding gauge field by the Kaluza--Klein reduction \cite{Morrison:1996pp, Dasgupta:1999ss}. 

For this, we 
explicitly compute the monodromy of two-cycles over the base $S$ which is related to the Poincar\'e dual to the certain topological $(1,1)$--forms. The $U(1)$ symmetry associated to the monodromy invariant two-cycle does not belong to the Cartan subalgebra of a non-Abelian gauge group which can be realized by a singularity of the Calabi--Yau fourfolds.  
Investigation of monodromy invariant two-cycles in F-theory has been extensively studied in \cite{Hayashi:2010zp}. Applying this technique, we can show that the monodromy for two-cycles is indeed reduced under the decomposition conditions predicted from the heterotic string analysis. In Section 3, we will consider an $E_6$ gauge theory with $(1+2)$ decomposition as an example, and find that the monodromy group of the two-cycles is reduced at least in a subspace of a moduli space in a stable degeneration limit of the Calabi--Yau fourfolds. Furthermore, we will see that the monodromy locus for $SU(5)$ gauge theories is globally factorized and contains squared factors under the decomposition conditions, which may provide an evidence that the monodromy group is reduced. 
In practice, this provides a nice way of constructing a $U(1)$ gauge symmetry on the F-theory side by the geometric data of the Calabi--Yau fourfolds. Without the prediction from heterotic string theory, it would not be easy to guess under which condition we may have $U(1)$ gauge symmetries. By bottom-up construction, we first specify the geometry and identify singularities to design the model reproducing the gauge theory that we want \cite{bottomup}. Our work would have a potential for the application to design a $U(1)$ symmetry in F-theory compactifications. 

There are also another resolutions to obtain a $U(1)$ symmetry in F-theory compactifications by other approaches \cite{globalu1, Grimm:2010ez}. Massive Abelian gauge symmetries associated to non-harmonic forms are discussed in \cite{Grimm:2010ez, Grimm:2011tb}. Alternatively, a discrete symmetry can be used for phenomenological reasons such as the prohibition of the dimension-four proton decay operators. The geometric condition for it has been explored in \cite{Tatar:2009jk, Hayashi:2009bt}.

\section{The construction of decomposed spectral covers}
\label{sec:spectral}

After reviewing the spectral cover construction for a vector bundle with an $SU(5)$ structure group, 
we will explicitly construct decomposed spectral covers for bundles whose structure groups are $S(U(1) \times U(4))$, $S(U(2) \times U(3))$ and $S(U(1) \times U(1) \times U(3))$. In the construction, we make use of the group 
of points on an elliptic curve. Basic facts of the group 
on an elliptic curve is summarized in Appendix \ref{sec:groupsum}.

\subsection{Spectral cover for $SU(n)$} \label{sec:sun}

Consider $E_8 \times E_8$ heterotic string compactified on a three dimensional complex manifold $Z_3$, from which we want to obtain a realistic model in four real dimensions. For a supersymmetric vacuum, the requirements for the vanishing of the variations of fermionic fields give us the following conditions \cite{Candelas:1985en};
\begin{itemize}
\item We need chiral gauge theory in four dimensions, which is possible with ${\cal N}=1$ supersymmetry. From the transformation of gravitino, $Z_3$ should have an $SU(3)$ holonomy. The required $Z_3$ is a Calabi--Yau threefold.
\item For gaugino, the background gauge field on $Z_3$ should satisfy the Hermitian Yang--Mills (HYM) equations. 
 This condition is satisfied by a vector bundle having the property called stability \cite{D, UY}. If this vector bundle $V$ has a structure group $G$, the low energy effective field theory is a gauge theory with an unbroken group as the commutant of $G$ in $E_8$.
\end{itemize}

A good thing is, if $Z_3$ admits an elliptic fibration over the base $S$ with a section, we can explicitly construct a stable vector bundle of the structure group $G$ \cite{Friedman:1997yq, Donagi:1997dp, Friedman:1998si}. We use the Weierstrass equation for the elliptic fiber $E$
\begin{equation} \label{Wei}
 y^2 = x^3 + f x + g, \quad x,y \in \C.
\end{equation}
From the Calabi--Yau condition, $f$ and $g$ are global holomorphic sections, $f \in \Gamma(S; \cO(-4K_S))$, $g \in \Gamma(S; \cO(-6K_S))$ respectively. Here, $K_S$ denotes the canonical bundle of $S$.
We have a section at the `origin' or `point at infinity' $o:(x,y)\to(\infty,\infty)$, whose meaning is clear if we embed this into a projective space; see Appendix \ref{sec:groupsum}. Let us first consider a vector bundle $V$ with a structure group $SU(n)$. 
A semistable vector bundle $V$ on a single elliptic curve $E_p$ at a point $p \in S$ can be derived from a set of degree zero line bundles $N_i$, trivially satisfying the HYM equations \cite{Borel}. 
Every degree zero line bundle $N_i$ on $E_p$ is related to a unique point $Q_i$ on $E$, such that $N_i$ has a holomorphic section with a zero at $Q_i$ and a pole at $o$ (see for example \cite{Shafarevich}). In other words, on $E_p$ there is a one-to-one correspondence between
$$ \text{a degree zero line bundle }N_i  \ \Longleftrightarrow\ \text{a divisor } Q_i-o.$$
In particular $Q_i = o$ parameterizes a trivial line bundle.
On a single elliptic curve $E_p$, a set of $n$ points in reference to an origin $o$ on $E_p$ parameterizes $n$ copies of such line bundles, whose direct sum will form a vector bundle of $SU(n)$ ($SU(n)$ not $U(n)$ if the tensor products of the line bundles is a trivial line bundle). As we fiber $E_p$ over the base twofold $Z_3 \to S$ to make $E$, this set of points $Q_i$'s become $n$-fold cover of $S$ with a possible twist, which is called a spectral cover $C_V$. When a line bundle ${\cal N}_V$ on $C_V$ is given, a stable $SU(n)$ vector bundle can be obtained by the Fourier-Mukai transform of the spectral data $(C_V, {\cal N}_V)$ \cite{Friedman:1997yq}. Via fibration the above direct sum of line bundles is deformed to a nontrivial stable vector bundle under the structure group $SU(n)$.

For concreteness, we first consider an $SU(5)$ spectral cover. We need to specify five points $Q_i,i=1,2,3,4,5$ as the locations of zeros of an equation. For the structure group is unimodular, the determinant of the vector bundle is trivial and the tensor product is related to the group sum of the torus $E_p$ such that\footnote{In this paper we will denote the group sum and subtraction by respectively $\boxplus$ and $\boxminus$. 
}
\begin{equation}  \label{zerosum}
 Q_1 \boxplus Q_2 \boxplus Q_3 \boxplus Q_4 \boxplus Q_5= o .
\end{equation}
These are specified by the zeros of a following function
\begin{equation} \label{su5cover}
w(x,y) = a_0 + a_2 x + a_3 y + a_4 x^2 + a_5 xy,
\end{equation}
where $x$ and $y$ are the coordinates on $E$, as in (\ref{Wei}). We can see how Eq.~\eqref{su5cover} specifies five points $Q_i \; (i=1, \cdots, 5)$ as follows. One can eliminate the dependence of $y$ in \eqref{su5cover} by using \eqref{Wei}. Then, Eq.~\eqref{su5cover} becomes a degree five equation in $x$. The root of the equation specifies the $x$ coordinate of five $Q_i$'s. Inserting the $x$ coordinates into \eqref{su5cover}, we can also get the $y$ coordinate of five $Q_i$'s. Hence, Eq.~\eqref{su5cover} specifies five points $Q_i$'s on $E$. Including $o$, Eq.~\eqref{su5cover} can be viewed as a meromorphic function with an order five pole at $o$.  This means, the points satisfy the constraint (\ref{zerosum}), due to the theorem that the sum of zeros is equal to sum of poles (see, for example \cite{Hurwitz}).

In the special case
\begin{equation} \label{tencurve}
  a_5=0,
\end{equation}
the spectral cover (\ref{su5cover}) is reduced to that of $SU(4)$. We have gauge symmetry enhancement of the commutant group from $SU(5)$ to $SO(10)$ in $E_8$. Then, one root of the spectral surface equation $w(x, y)=0$ is always at the point at infinity $o$, and it indicates the presence of massless matter fields in the representation of ${\bf 10}$ or ${\bf \overline{10}}$ under the unbroken $SU(5)$ gauge symmetry. On the other hand, the dual F-theory picture shows clearly that, from the branching of $SO(10)$ adjoint $\bf 45 \to 24 + 10 + \overline{10} + 1$, we have matter fields having quantum number $\bf 10$ or $\overline{{\bf 10}}$ of $SU(5)$, 
and the ${\bf 10} + {\bf \overline{10}}$ matter fields are localized on the curve given by \eqref{tencurve} on $S$ \cite{Katz:1996xe}.

Another case is when
 $Q_1 \boxplus Q_2 = o$ and $Q_3 \boxplus Q_4 \boxplus Q_5=o$, the structure group decomposes into $SU(5) \to SU(2) \times SU(3)$. So the unbroken group becomes larger group $SU(6)$ which is the commutant in $E_8$. In this case, a root  $Q_1 \boxplus Q_2$ corresponding to a weight of the ${\bf 10}$ representation of the structure group $SU(5)$ is always at the point at infinity $o$. This also indicates the presence of the massless matter fields in the representation of (${\bf \overline{5}} + {\bf 5}$) under the unbroken $SU(5)$ gauge group. At this locus Equation (\ref{su5cover}) takes a form
\begin{equation} \label{23factor}
 w(x,y) \to (A+Bx)(C+Dx+Ey)
\end{equation}
and the factorization condition is translated to
\begin{equation} \label{fivecurve}
 a_0 a_5^2 - a_2 a_3 a_5 + a_3^2 a_4 = 0,
\end{equation}
where $\bf 5 + \overline 5$ matter fields are localized in a similar but nontrivial way \cite{Hayashi:2008ba, Donagi:2009ra}.

These two cases exhaust the possible matter curves by gauge symmetry enhancement of $SU(5)$. In the next section we will see that it is not possible to take a factorization including $U(1)$ as in (\ref{23factor}). 

Generalizing we can construct spectral cover equations containing terms upto $x^3$, which is always linear in $y$ with the aid of (\ref{Wei}). This spectral cover construction applies for the structure groups of $SU, Sp$ types. Other vector bundles with $SO$ type and exceptional structure groups can be constructed by different method \cite{Friedman:1997yq}.

\subsection{Parameterizing $U(1)$ by a globally isolated point}
\label{sec:uone}

A $U(1)$ line bundle is parameterized by one point, and it is nontrivial if the point is away from the reference point $o$. Although there is no obstruction to put a point $Q$ away from $o$, it is not possible to describe such a point by an equation like (\ref{su5cover}), which would be a meromorphic function having an order one pole only at $o$ if it exists. This is due to a theorem that there is no elliptic function with a order one pole \cite{Hurwitz}. 
Moreover, there is no preferred choice of an origin $o$ on the torus. So it is meaningless to say a point away from $o$ one if we consider one point.

These problems do not arise if we consider $U(1)$ or $U(n)$ as a subgroup of the structure group $SU(m)$ with $n<m$. The former can be obtained by tuning of the parameters of the spectral cover describing the latter.
For concreteness, consider a `decomposed' structure group as a subgroup of a simple group
$$S(U(1)\times U(4)) \subset SU(5).$$
Being it the subgroup of $SU(5)$, we have five points as zeros of the spectral cover $w$ in (\ref{su5cover}) which satisfy the zero sum relation (\ref{zerosum}). Then we partition five points into two sets, say, $\{ Q_1\}$ and $\{Q_2,Q_3,Q_4,Q_5\}$ which are 
connected by 
the monodromy 
only on the latter.
Since they are still embedded in $SU(5)$, we have
\begin{equation}
 \label{u4cond} Q_2 \boxplus Q_3 \boxplus Q_4 \boxplus Q_5 = \boxminus Q_1 \ne o.
\end{equation}
For this purpose, the sufficient conditions for such a point $Q_1$ is that the location of the $Q_1$ should be
\begin{enumerate}
\item {\em globally} isolated, otherwise we may still have 
monodromy mixing all the five when we move over the base $S$.
\item different from $o$, otherwise $Q_1$ will parameterize trivial bundle.
\end{enumerate}
So our elliptic fiber should admit a more global section describing $Q_1$. This means, the coordinate of $Q_1$, say $(x_1,y_1)$, parameterizes the global section. Since it is on $E$ we have from (\ref{Wei})
\begin{equation} \label{gsol}
 y_1^2 = x_1^3 + f x_1 + g.
\end{equation}
Since $(x_1, y_1)$ parameterizes a global section, Equation \eqref{gsol} should be satisfied \emph{automatically} regardless of $f$ and $g$. In other words, \eqref{gsol} should not be regarded as a constraining equation for $x_1$ and $y_1$. 
Hence, this tunes $E$ by fixing one of the parameters $f$ or $g$ in terms of $(x_1,y_1)$.
Solving $g$ in terms of others, and plugging it to the original equation (\ref{Wei}) yields
\begin{equation} \label{Wei2sec}
 (y-y_1)(y+ y_1) = (x-x_1)(x^2 + x_1 x+ x_1^2 + f).
\end{equation}
This factorization structure persists even if we move over the base $S$.
Other {\em generic} points on $E$ cannot have the coordinates as definite global sections, having the factorization as in (\ref{Wei2sec}) (but we may see some special point such as $Q_1 \boxplus Q_1$ can be such a global point). Because of the Weierstrass form (\ref{Wei2sec}), we always have {\em another} global section at $\boxminus Q_1 :(x_1,-y_1)$ as well. 
If we further require this $Q_1$ be on the $SU(5)$ spectral cover, its coordinates $(x_1,y_1)$ should satisfy Equation (\ref{su5cover})
\begin{equation} \label{u1tuning}
 w(x_1,y_1)=0.
\end{equation}
This condition may constrain $a_r$'s. A simple solution of \eqref{u1tuning} is 
\be
a_0 = -a_2 x_1 - a_3 y_1 - a_4 x_1^2 - a_5 x_1 y_1.
\label{eq:simplesol}
\ee
In fact, one may also obtain further conditions for $a_r$'s from the compatibility of the two equations \eqref{gsol} and \eqref{u1tuning} like the {\rm holomorphy} of $f$ or $g$. We will see these conditions in Section \ref{sec:Hetcase1} and \ref{sec:Hetcase2}.
These two conditions (\ref{gsol}) and (\ref{u1tuning}) are sufficient to guarantee the presence of a globally isolated point $Q_1$ describing $U(1)$.

A comment on stability involving $U(1)$ or $U(n)$ structure group. The Refs.~\cite{D, UY} used a stability condition using a quantity called $\mu$-slope of a vector bundle. The $\mu$-slope for a bundle $V$ is defined as 
\be
\mu(V) = \frac{1}{{{\rm rk}(V)}} \int_{Z_3}c_1(V) \wedge J \wedge J,
\ee
where $J$ is the K\"ahler form on $Z_3$. A vector bundle $V$ is stable if there is no subbundle $F$ of a lower rank structure group whose slope satisfies $\mu(F) < \mu(V)$. By this definition $U(1)$ is stable. Even if an $SU(n+1)$ bundle $V$ were reducible, for instance, a direct sum of a line bundle $L$ and a rank $n$ vector bundle $V^{\prime}$ with a structure group $S(U(1) \times U(n))$, it could still satisfy the HYM equation, if the $\mu$-slopes of irreducible pieces are all equal $\mu(L) = \mu(V^{\prime}) = \mu(V)$ and each of them is stable \cite{D, UY}. In this case the bundle $V$ is called {\em poly-stable}. Then, the $S(U(1) \times U(n))$ vector bundle constructed from a decomposed spectral cover could be poly-stable if $\mu(L) = \mu(V^{\prime}) = 0$. This is because we have $\mu(V) = 0$ and $\mu(L) = -\mu(V^{\prime})$ from the specialty condition of the $SU(n+1)$ bundle. Note that $c_1(L)=0$ nor $c_1(V^{\prime}) = 0$ are not required to satisfy $\mu(L) = \mu(V^{\prime}) = 0$ generically. This stability story has a clear physical picture. In fact, $\mu(L)$ is related a Fayet-Illiopoulos term, and the non-zero value for $\mu(L)$ can give rise to a VEV to bundle moduli \cite{Blumenhagen:2005ga, Tatar:2006dc}. By the certain signs of the Fayet-Illiopoulos term and $U(1)$ charged fields, the reducible bundle $V$ becomes an extension of $V^{\prime}$ by $L$ \cite{Friedman:1997yq, Tatar:2006dc}, namely $V$ becomes rank $5$ irreducible bundle. 

\subsection{Embedding to $Sp(n)$ spectral cover}

For explicit visualization of the decomposition $S(U(1) \times U(4)) \subset SU(5)$, it is useful to embed the $SU(5)$ spectral cover into an $Sp(5)$ cover. We may calculate the actual coordinates of $Q_i:(x_i,y_i)$. Equating $w=0$ we have
\begin{equation} \label{weq}
 a_0 + a_2 x + a_4 x^2 = -(a_3 + a_5 x)y.
\end{equation}
Squaring it and plugging (\ref{Wei}), we have
\begin{equation}
 (a_0 + a_2 x + a_4 x^2)^2 = (a_3 + a_5 x)^2 y^2 =(a_3 + a_5 x)^2 (x^3+f x+g),
\end{equation}
or,
\begin{equation}  \label{deg5eq}
\begin{split}
 a_5^2 x^5 &+ ( 2 a_3 a_5-a_4^2 ) x^4+ (a_3^2-2 a_2 a_4+a_5^2 f) x^3 + (-a_2^2 - 2 a_0 a_4 + 2 a_3 a_5 f + a_5^2 g) x^2 \\
  &+(-2 a_0 a_2 + a_3^2 f + 2 a_3 a_5 g) x+ (a_3^2 g-a_0^2) =0.
 \end{split}
\end{equation}
Being a degree five polynomial of single variable over the complete field $\C$, it has five roots $x_i$. Putting them back to (\ref{weq}), we obtain the rest of the coordinates $y_i$. Using the group relations on $E_p$, (\ref{sumx}) and (\ref{sumy}) in Appendix \ref{sec:groupsum}, for the coordinates $Q_i:(x_i,y_i)$, we can verify the zero sum relation (\ref{zerosum}) required for the unimodular group $SU(5)$. This is reflected that our polynomial in (\ref{deg5eq}) is non-generic but has highly nontrivially tuned coefficients .

Without (\ref{weq}), the single equation (\ref{deg5eq}) has the following properties.  It has symmetry under $y \to -y$, meaning that, for each root for $Q_i:(x_i,y_i)$, we always also have a root for $\boxminus Q:(x_i,-y_i)$.
This means, its solution gives five pairs of points
\begin{equation} \label{sp5points}
 Q_1, Q_2, Q_3, Q_4, Q_5\text{ and } \boxminus Q_1, \boxminus Q_2, \boxminus Q_3, \boxminus Q_4, \boxminus Q_5.
 \end{equation}
With generic coefficients, a degree five polynomial in $x$ describes five pairs of points, comprising an $Sp(5)$ spectral cover. In our case, the coefficients are tuned such that, the $x$-coordinate of every zero $Q_i$ of the original $SU(5)$ cover polynomial (\ref{su5cover}) always satisfies our equation (\ref{deg5eq}). 

Since Eq.~\eqref{deg5eq} are written only by $x$ and $x$ is not constrained in the equation, the decomposition $S(U(1) \times U(4))$ should become clear if one considers \eqref{deg5eq}. Namely, Eq.~\eqref{deg5eq} must be factorized if one exploits the parameterization for the spectral cover of $S(U(1) \times U(4))$ vector bundle. Let us consider the implication of the presence of a global section at $Q_1$. Its coordinates $(x_1,y_1)$ automatically satisfy the elliptic equation (\ref{Wei2sec}).  Noting $Q_1:(x_1,y_1)$ is on the spectral cover, we have $w(x_1,y_1)=0$ from (\ref{su5cover}). Let us now apply a simple solution of \eqref{eq:simplesol}. Even if we eliminate $y$ from these two, still $x=x_1$ should be the solution to the equation. 
then, (\ref{deg5eq}) is factorized as
\begin{equation} \label{factorization}
 (x - x_1)(D_0 + D_1 x + D_2 x^2 + D_3 x^3+ D_4 x^4) = 0,
\end{equation}
where $D_0, D_1, D_2, D_3, D_4$ are functions of $x_1, y_1, a_2, a_3, a_4, a_5$ and $f$. 
For generic coefficients, the factors in (\ref{factorization}) parameterize the points related to minimal weights of $Sp(1)$ and $Sp(4)$, respectively. However, if they are tuned such that the expansion of (\ref{factorization}) has a form as (\ref{deg5eq}) and  $y$-coordinates satisfy the relation (\ref{su5cover}), we can say they describe a subgroup of the $SU(5)$ structure group. The common intersection is
$$ S(U(1)\times U(4)).$$
Therefore our remaining task is to find $D_i$'s in (\ref{factorization}).
Although it is hard to find the most general form, we find two useful particular solutions in the following.

Generalizing, we can embed an $SU(n)$ spectral cover into an $Sp(n)$ cover. In addition to (\ref{weq}) we may consider higher order terms in $x$ and $y$, in which we can always make the $y$-dependent linear using the Weierstrass equation (\ref{Wei}).

\subsection{$S(U(1) \times U(4))$ spectral cover - case I, a simple ansatz}
\label{sec:Hetcase1}

We {\em demand} the presence of another global section at a point $Q_1$ other than $o$.
In the special case in which the $x$-coordinate of $Q_1$ is 0, the condition of its being on $SU(5)$ spectral cover reads $w(0,y_1)=0$, or
$$ a_0 + a_3 y_1 = 0. $$
We combine this with the fact that $Q_1$ is a global section of the elliptic fibration $E$, then we have conditions
\begin{equation} \label{gcond}
 a_3^2 y_1^2 = a_3^2 g = a_0^3.
\end{equation}
To have well-defined $E$, $g$ must be global {\em holomorphic} sections over $S$. This means, from the relation (\ref{gcond}) we need a global holomorphic section $a_{-3}$ over $S$ such that
\begin{equation} \label{a3factor}
 a_3 a_{-3} = a_0, \qquad g = a_{-3}^2.
\end{equation}
Note that the condition for the complex structure moduli of $Z_3$ is inevitable in this case.
The elliptic equation (\ref{Wei}) becomes
\begin{equation} \label{moresections}
 \left(y+ a_{-3} \right ) \left( y- a_{-3} \right) =  x \left(x^2+ f \right),
\end{equation}
which explicitly shows that presence of two more global sections at $(x,y)=(0, \pm a_{-3})$ which are nothing but $Q_1$ and $\boxminus Q_1$. This is independent of $a_3$ so is valid for $a_3 \to 0$. In the sense that (\ref{moresections}) preserves the form when we mover around the base $S$, there is no monodromy mixing $Q_1$ with the other $Q_i$'s, as desired. The other values $x$ solving $x^2 + f = 0$ changes the sign as we move around $S$, so they cannot provide global sections.

Here, we have used a different logic from the one in Section \ref{sec:uone}. We first assumed that $x_1 = 0$ and used \eqref{u1tuning} to determine $y_1$, instead of constraining $a_r$'s. The factorization condition on $a_0$ in \eqref{a3factor}, in fact, have come from the {\em holomorphy} of $g$, or to make the global section well-defined on whole $S$.

Using the parameterization (\ref{a3factor}), our spectral cover equation becomes
\begin{equation}
 a_3 a_{-3} + a_2 x + a_3 y + a_4 x^2 + a_5 xy = 0. \label{14-1}
\end{equation}
The decomposition structure is not manifest at this stage, however embedding into the $Sp(5)$ spectral cover equation, we can see
\begin{equation} \begin{split}
 \ x  \big[& a_5^2 x^4+(2 a_3 a_5 - a_4^2) x^3  +  (a_3^2 - 2 a_2 a_4 + a_5^2 f )x^2  \\
   &+ (a_5^2 a_{-3}^2 - a_2^2 - 2 a_3 a_4 a_{-3} + 2 a_3 a_5 f)x  + 2 a_3 a_5 a_{-3}^2 - 2 a_2 a_3 a_{-3} + a_3^2 f \big] = 0,
\end{split}
\end{equation}
which is in the form (\ref{factorization}). 

Since we have $1+4$ partitions of points hence small monodromy group
, some of the matter curves are distinguished. We expect an additional $U(1)$ gauge or global symmetry in the low energy effective theory is also realized. The latter is a case when the corresponding gauge boson becomes massive by the Green--Schwarz mechanism if the $U(1)$ symmetry is anomalous. 

The commutant to $\S(U(1)\times U(4))$ in $E_8$ is $SU(5)\times U(1)$. In terms of group theory, we can distinguish the representations of $SU(5)$ which are non-trivially charged under the $U(1)$. The gauge enhancement loci (\ref{tencurve}) and (\ref{fivecurve}) associated with the matter of the unbroken $SU(5) \times U(1)$ gauge group become
\begin{align}
   SO(10) \times U(1) &:a_5 = 0 , \label{so10enh} \\
  SU(6) \times U(1) &: a_3 = 0 \\
  SU(6) \times U(1) &:  a_{-3} a_5^2 - a_2 a_5 + a_3 a_4 = 0, \label{su6enh}
\end{align}
under the parameterization (\ref{a3factor}). The relation (\ref{fivecurve}) is factorized, so the vanishing of each factor give rise to $SU(6) \times U(1)$ symmetry. When $a_3 = 0$, then the spectral cover equation \eqref{14-1} becomes \be
x(a_2 + a_4 x + a_5 y) = 0. \label{a3zero}
\ee
The first factor of \eqref{a3zero} is a spectral cover for a $S(U(1) \times U(1))$ bundle since $x=0$ in the elliptic fiber $E$ corresponds to two global sections $Q_1$ and $\boxminus Q_1$. The second factor of \eqref{a3zero} describes a spectral cover for an $SU(3)$ bundle. Hence the structure group on $a_3 = 0$ is $S(U(1) \times U(1)) \times SU(3)$, and the unbroken gauge group is $SU(6) \times U(1)$. On the other hand, the structure group on \eqref{su6enh} is $SU(2) \times S(U(1) \times U(2))$ since the $(C + Dx + Ey)$ factor of \eqref{14-1} satisfies $C + D \cdot 0 + E (a_{-3}) = 0$ on \eqref{su6enh}. Then the unbroken gauge group on the $\overline{{\bf 5}}$ matter curve \eqref{su6enh} becomes $SU(6) \times U(1)$.    
Without demanding the holomorphy of $g$, resulted in the structure (\ref{a3factor}), the matter curve would not be factorized, not reflecting this fact. 
The relation (\ref{tencurve}) is unaffected since $Q_1$ never hits $o$.

The moduli space for the $S(U(1)\times U(4))$ vector bundle on $Z_3$ can be characterized by $a_2, a_3, a_4, a_5, a_{-3}$.  
In fact, $a_{-3}$ participates in the spectral surface equation \eqref{14-1} and also the defining equation of $Z_3$. So, $a_{-3}$ may be regarded as the complex structure moduli of $Z_3$ as well as the vector bundle moduli. 
In the previous case of $SU(n)$, discussed in Section \ref{sec:sun}, the complex structure moduli of $Z_3$ and the vector bundle moduli were not affecting each other. In this case we cannot separate the moduli space of the complex structure of $Z_3$ from that of the vector bundle, since the description of the $U(1)$ cover is determined by the location of the global section at $Q_1$ different from $o$. 

\subsection{$S(U(1) \times U(4))$ spectral cover - case II, Higgs bundle analogy}
\label{sec:Hetcase2}

For generic $f$ and $g$, it would not be possible to decompose the embedded $SU(5)$ equation (\ref{deg5eq}) into an $1+4$ form (\ref{factorization}). Nevertheless, we have a fairly general class of solutions, hinted by the Higgs bundle picture in F-theory compactification \cite{Hayashi:2009ge, Donagi:2009ra}. There, the divisors $\{Q_i - o\}_{i=1,\dots,5}$ are related to VEV's of the adjoint Higgs scalar on 7-branes on the F-theory side. The latter is also related to the weights $\{t_i\}_{i=1,\dots,5}$ of the fundamental representation $\bf 5$ of the structure group $SU(5)$. Since we also have $S_5$ symmetry under the Weyl group, faithful representations come from elementary symmetric polynomials $s_k$ of orders $k$ in $t_i$,
\begin{equation} \label{grouprel}
\frac{a_k}{a_0} \sim s_k , \quad \prod_{i=1}^5(t+t_i) = \sum_{k=0}^5 s_k t^{5-k},
\end{equation}
with $a_0 \ne 0$.

The equation (\ref{grouprel}) can be captured by the spectral surface equation (\ref{su5cover}) in heterotic string theory in the following way. In the vicinity of the zero section $o$,
the elliptic functions $x$ and $y$ behave like $x \simeq s^{-2}, y \simeq s^{-3}$, where $s$ denotes a complex number of the Weierstrass $p$-function on the torus and $s=0$ is the point at infinity. Then the equation (\ref{su5cover}) {\em approximately} becomes
$$ a_0 s^5 + a_2 s^3 + a_3 s^2 + a_4 s + a_5 = 0.   $$
In this  case, factorization is possible in a form 
\begin{equation} \label{approxfactor}
 (b_0 s+b_1)(d_0 s^4+d_1 s^3+d_2 s^2 + d_3 s+d_4 ) = 0.
\end{equation}
This factorization means the coefficients $a_i$'s are tuned as follows 
\begin{align}
 a_0 &= b_0 d_0 , \label{14-0}\\
   0 &= b_0 d_1 + b_1 d_0 , \label{a1cond} \\
 a_2 &= b_0 d_2 + b_1 d_1, \\
 a_3 &= b_0 d_3 + b_1 d_2, \\
 a_4 &= b_0 d_4 + b_1 d_3, \\
 a_5 &= b_1 d_4. \label{14-5}
\end{align}
The condition (\ref{a1cond}) is understood as a traceless condition of the whole group. We also assume that $b_0$ and $b_1$ do not simultaneously vanish at any point in $S$ otherwise the spectral cover equation \eqref{su5cover} vanish under the conditions \eqref{14-0}--\eqref{14-5} at the point. It turns out that $d_0$ requires further tuning.

Coming back to exact parameterization, we {\em try} this parameterization and see what happens. Since $x \simeq s^{-2}$, the zero at $s = -b_1/b_0$ in the first factor is only mimicked by employing
\begin{equation} \label{sp1factor}
 b_1^2 x - b_0^2,
\end{equation}
of which $Q_1$ and $\boxminus Q_1$ are two zeros, with $x_1=b_0^2/b_1^2$. Putting this in the spectral cover equation (\ref{su5cover}) we find $y=-b_0^3/b_1^3$, agreeing with the approximation $y \simeq s^{-3}$. This yields a condition $b_0^2 f+b_1^2 g = 0$. On top of this we need further tuning of some parameters for the holomorphy. From $b_0^2 f+b_1^2 g = 0$
we require a holomorphic section $F$ such that $f=b_1^2 F$, which fixes $g=-b_0^2 F$ as well. From the relation (\ref{a1cond}), either $b_1$ or $d_0$ should be divisible by $b_0$. The former rather gives a trivial factor in (\ref{sp1factor}) we choose a holomorphic section $d$ such that $d_0=b_0 d$ fixing $d_1=-b_1 d$.
To sum up, we choose extra tuning for holomorphy
\begin{equation} \label{holtuning}
 f=b_1^2 F, \quad g=-b_0^2 F, \quad d_0 = b_0 d, \quad d_1 = -b_1 d.
\end{equation}
Hence, the parameterization of the $S(U(1) \times U(4))$ spectral surface is \eqref{14-0}--\eqref{14-5} with \eqref{holtuning}. Then we can show that (\ref{deg5eq}) is factorized as
\begin{equation} \label{14factor} \begin{split}
  \Big[ b_1^2 x-b_0^2 \Big]  &\Big[     d_4^2 x^4 + ( 2  d_2 d_4- d_3^2 ) x^3
    + ( d_2^2 + 2 b_1 d d_3 + 2  b_0 d d_4 +   b_1^2 d_4^2 F) x^2 \\
    &+( 2  d_0 d_2 -b_1^2 d^2 + 2 b_1^2 d_2 d_4 F +     2 b_0 b_1 d_3 d_4 F) x \\
    &+ b_0^2 d^2 +  b_1^2 d_2^2 F  +2 b_0 b_1 d_2 d_3 F  +  b_0^2 d_3^2 F \Big] = 0
    \end{split}
\end{equation}
Now the second factor should describe the four points $\{ Q_2, Q_3, Q_4, Q_5 \}$. This is a degree four polynomial in $x$ over $\C$, having four roots. Plugging them to spectral cover equation (\ref{su5cover}), we obtain the desired coordinates of the four points satisfying the constraint (\ref{u4cond}).

The second factor in (\ref{14factor}) is a $U(4)$ part. We have apparent dependence on $b_0$ and $b_1$ since they are more primitive than $d_0$ and $d_1$, from (\ref{holtuning}). If we wish to express only in terms of $d_0$ and $d_1$, a stronger condition $d=1$ is necessary.
We can obtain a similar equation from the $SU(4)$ spectral cover equation $b_0 + b_2 x + b_3 y + b_4 x^2=0$. Compared with this, our $U(4)$ factor contains additional terms 
$$ b_1\big[(2 d d_3 + b_1 d_4^2 F) x^2 + (-b_1 d^2 + 2 b_1 d_2 d_4 F + 2 b_0  d_3 d_4 F) x + (b_1 d_2^2 F+2 b_0  d_2 d_3 F)\big], $$
all proportional to $b_1$ hence to $d_1$.  Viewed as tuning parameters of the original spectral cover equation (\ref{su5cover}), we could not eliminate the dependence on $b_0$ and $b_1$ without ruining holomorphy. In this set of solutions, the group theoretical relations in the Higgs bundle picture like (\ref{grouprel}) are suggestive.  In particular $Q_2 \boxplus Q_3 \boxplus Q_4 \boxplus Q_5$ is related to $d_1$ hence to $b_1$ which is related to $Q_1$. It is also understood that the first $U(1)$ factor is described by $b_1/b_0$ whereas another overall $U(1)$ as a part of $U(4)$ is $-d_1/d_0$. 

The gauge symmetry enhancement conditions (\ref{tencurve}) and (\ref{fivecurve}) further factorize as follows
\begin{align}
 {\bf \overline 5 }&:   (b_1^2 d_2 + b_0 b_1 d_3 + b_0^2 d_4) ( b_1 d_2 d_3 + b_0 d_3^2 +
   b_1^2 d d_4) = 0 \\
  {\bf 10}&:  b_1 d_4 = 0
\end{align}
In the special case $b_0=1$ we recover $d = d_0$, which have been well-known form in the Higgs bundle picture.
In the F-theory side, the factorization is reflected by a monodromy locus. Even if we consider higher order terms of the defining equation of geometry, we will explicitly see in the next section the reduction of monodromy at least in a subspace in a moduli space and the possibility of the appearance of a new monodromy invariant two-cycle which harbors the new Cartan algebra.

We have obtained the coordinate of $Q_1$ as $(x_1,y_1)=(b_0^2/b_1^2,-b_0^3/b_1^3)$. This is valid for $b_1 \ne 0$ but the limit $ b_1 \to 0 $ is understood that $Q_1$ is going to $o$. We now check that $(x_1, y_1) = (b_0^2/b_1^2,-b_0^3/b_1^3)$ is indeed a global section of the elliptic fibration $E$ under the parameterization \eqref{holtuning}.
Going to homogeneous coordinate $[X,Y,Z]$ in $\P^2$, we can regard the equation (\ref{Wei2sec}) as affine form at the patch $Z = 1$ and $o$ corresponds to $Z=0$. Multiplying $b_1^6$ to the homogeneous equation, we have
\begin{equation} \label{b1eq}
 (b_1^3 Y - b_0^3 Z)(b_1^3 Y+b_0^3 Z)Z - (b_1^2 X-b_0^2 Z)(b_1^4 X^2 + b_1^2 b_0^2 X Z +b_0^4 Z^2 +  b_1^6 F Z^2) = 0.
\end{equation}
Since the equation \eqref{b1eq} explicitly has the factors $(b_1^3 Y + b_0^3 Z)$ and $(b_1^2 X - b_0^2 Z)$, and hence $(b_1^3 Y + b_0^3 Z) = (b_1^2 X - b_0^2 Z) = 0$ is always a solution of \eqref{b1eq}. This means that $(x_1, y_1) = (b_0^2/b_1^2,-b_0^3/b_1^3)$ is a global section of the elliptic fibration $E$. Note that the equation \eqref{b1eq} is only valid if $b_1 \neq 0$. When $b_1 = 0$ but $b_0 \neq 0$, then the section $(b_1^3 Y + b_0^3 Z) = (b_1^2 X - b_0^2 Z) = 0$ specifies a point $Z = 0$. Hence it is still a section in a region where $b_1 = 0$ but $b_0 \neq 0$. Note that we have assumed that there is no region where $b_1$ and $b_0$ simultaneously vanish at any point in $S$ in order that the Higgs bundle ansatze \eqref{14-0}--\eqref{14-5} define a well-defined spectral cover equation on whole $S$. Hence, we have shown that $(b_1^3 Y + b_0^3 Z) = (b_1^2 X - b_0^2 Z) = 0$ is a global section of the elliptic fibration all over $S$.


With the decomposition conditions \eqref{14-0}--\eqref{14-5} and \eqref{holtuning}, we have global holomorphic sections $b_0, b_1, d, d_2, d_3, d_4$ and $F$. $b_0$ and $b_1$ appear in both the $S(U(1) \times U(4))$ spectral cover equation and also the defining equation \eqref{Wei} of $Z_3$. So, the complex parameters in the sections $b_0$ and $b_1$ may be regarded as the complex structure moduli as well as the bundle moduli. This characteristic was also observed in the case I in Section \ref{sec:Hetcase1}. There is no clear separation between the complex structure moduli space of $Z_3$ and the bundle moduli space.

Note that the tuning of the case II \eqref{14-0}--\eqref{14-5} with \eqref{holtuning} does not include the tuning of the case I \eqref{a3factor}. Since $g = a_{-3}^2$ from  \eqref{a3factor}, $F$ should be a section of a trivial line bundle from the relation $g =-b_1^2 F$. Certainly, $F = F_1^2$ can satisfy the relation $g = a_{-3}^2$ but this is a special case of $g = a_{-3}^2$. If $F$ is a section of a trivial line bundle, then $f \propto b_1^2$. However this ruins the genericity for $f$ of the case I. Therefore, the parameterization of the case II does not include the generic parameterization of the case I.



\subsection{$S(U(2)\times U(3))$ spectral cover}

Thus far we have described the decomposition of the spectral covers containing a $U(1)$ group, and we may also consider one containing $U(n)$. Here, we deal with a case 
\begin{equation} \label{su2u3}
 S(U(2)\times U(3)),
\end{equation}
 and the generalization is straightforward. 
 
Since (\ref{su2u3}) is a subgroup of $SU(5)$, the spectral cover is described by five points $\{Q_i\}$, satisfying the zero sum relation (\ref{zerosum}). We segregate these into two and three points 
$$ Q_1 \boxplus Q_2 = \boxminus (Q_3 \boxplus Q_4 \boxplus Q_5). $$
To reduce the monodromy, 
 each side of this equation should be the location of a global section but different from $o$. This amounts to introducing a point $Q$ as the global section, providing a `spectator' $U(1)$ group, and we can use the method discussed so far by considering $S(U(1)\times U(2))$ and $S(U(1)\times U(3))$ such that
\begin{equation} \label{23seg}
 Q_1 \boxplus Q_2 =  Q , \quad  Q_3 \boxplus Q_4 \boxplus Q_5 = \boxminus Q. 
\end{equation}

Let the explicit coordinates of $Q$ be
\begin{equation} 
 Q: \left(\frac{e_0^2}{e_1^2},\frac{e_0^3}{e_1^3} \right),
\end{equation}
in which $Q$ goes $o$ if $e_1 \to 0$ as before.
Then the Higgs bundle ansatze  \eqref{14-0}--\eqref{14-5} and \eqref{holtuning} become
\begin{align}
 S(U(1)\times U(2)):\ & e_0 b_0 + (e_0 b_2 - e_1 b_1)x -  e_1 b_2 y=0, \label{su1u2cover} \\
  & - e_0 b_1+e_1 b_0 = 0, \label{su1u2unimod} \\
 S(U(1)\times U(3)):\ & e_0 d_0 + (e_0 d_2 + e_1 d_1)x + (e_0 d_3 + e_1 d_2)y + e_1 d_3 x^2=0, \\
  &e_1 d_0+e_0 d_1=0, \label{su1u3unimod}
\end{align}
which are special cases of the above. Here we introduced $-b_1$ instead of $b_1$ without loss of generality for the reason that will be clear soon. Due to the relative signs of $Q$ and $\boxminus Q$ in (\ref{23seg}), (\ref{su1u2cover}) has an opposite sign of $y$.
As before, we need conditions for the holomorphy of the parameters in the elliptic equation (\ref{Wei}) and the solution to the constraints (\ref{su1u2unimod}) and (\ref{su1u3unimod}). We introduce global holomorphic sections $d$ and $F$ such that
\begin{equation} \label{su2u3hol}
 f= e_1^2F, \quad g = -e_0^2 F,\quad b_0 = e_0b, \quad b_1 =  e_1 b, \quad  d_0 = e_0 d,\quad d_1 = - e_1 d. 
\end{equation}
Note that $e_0$ and $e_1$ are more primitive than $b_0,b_1,d_0,d_1$, since the former are used for specifying the location of the global section that is necessary even for $U(2)$ or $U(3)$ factor.
Removing $y$ using the elliptic equation (\ref{Wei}), the resulting embedded equations $S(U(1)\times U(2)) \subset Sp(1) \times Sp(2)$ and
$S(U(1)\times U(3)) \subset Sp(1)\times Sp(3)$ have a common $U(1) \subset Sp(1)$ factor
$$e_1^2 x - e_0^2.$$
One can also show that $Q: \left(\frac{e_0^2}{e_1^2},\frac{e_0^3}{e_1^3} \right)$ parameterizes a global section of the elliptic fibration $E$ under the parametrization of \eqref{su2u3hol}.

From the two embedding equations $S(U(1)\times U(2)) \subset Sp(1) \times Sp(2)$ and
$S(U(1)\times U(3)) \subset Sp(1)\times Sp(3)$, one can extract the $U(2)$ factor and the $U(3)$ factor. The final factorization form of the embedded equation in $Sp(2) \times Sp(3)$ should be the product of the $U(2)$ factor and the $U(3)$ factor
\begin{equation}
\begin{split}
\big[& b_2^2 x^2+( 2 e_0 b b_2 - b^2 e_1^2)x+e_0^2 b^2 + e_1^2 b_2^2 F \big] \times \\
 & \big[
   d_3^2 x^3+(-   2 e_1 d d_3 -d_2^2 )x^2+ (e_1^2 d^2 -2 e_0 d d_2 )x
  - e_0^2 d^2 -e_1^2 d_2^2 F - 2 e_0 e_1 d_2 d_3 F - e_0^2 d_3^2 F
  \big]=0.
 \end{split} \label{23}
\end{equation}
Again, $e_0$ and $e_1$ are more primitive, so we cannot eliminate their dependence.

We need to find a parameterization for $a_r, (r=0, 2, 3, 4, 5)$ whose spectral surface equation becomes the form (\ref{23}) after eliminating $y$. Namely, we uniquely express $a_r$ in (\ref{deg5eq}), which are the coefficients of the $SU(5)$ spectral cover (\ref{su5cover}), in terms of $b_i$, $d_i$, $e_i$ and $F$. We find that
\begin{align}
a_0 &= b_0 d_0 +  e_1^2 b_2 d_2 F+ e_0 e_1 b_2 d_3 F , \label{23-0}\\
0  & = b_0 d_1 + b_1 d_0 , \label{traceless23} \\
a_2 &= b_0 d_2 + b_1 d_1 + b_2 d_0 , \\
a_3 &= b_0 d_3 + b_1 d_2 + b_2 d_1, \\
a_4 &= b_1 d_3 + b_2 d_2, \\
a_5 &= b_2 d_3. \label{23-5}
\end{align}
Here the relation (\ref{traceless23}) is automatic from the parameterization (\ref{su2u3hol}). Hence, the parameterization for the $S(U(2) \times U(3))$ spectral cover is \eqref{23-0}--\eqref{23-5} with \eqref{su2u3hol}. These relations are similar to what have been known in the Higgs bundle picture,
$$ (b_0s^2+b_1s+b_2)(d_0s^3 + d_1 s^2+d_2 s+ d_3) = 0 $$
except the dependence on $F$. The center-of-masses of $U(1),U(2),U(3)$ `branes' are respectively parameterized by $e_1/e_0,b_1/b_0,-d_1/d_0$ with none of $e_0,b_0,d_0$ being zero and they are related by the condition (\ref{su2u3hol}). 

\subsection{$S(U(1) \times U(1) \times U(3))$ spectral cover}

One can also construct a spectral cover which has more than one $U(1)$ cover. We have seen that Eq.~\eqref{Wei2sec} admits the global section at $Q$ and $\boxminus Q$ and $Q$ has been used for parameterizing the $U(1)$ cover of the $S(U(1) \times U(4))$ spectral cover. In fact, there are another global sections at $Q \boxplus Q$ and $\boxminus (Q \boxplus Q)$ also from the Weierstrass equation \eqref{Wei2sec}. The location of the global section at $Q \boxplus Q$ is 
\be
(x_2, y_2) = \left(\frac{b_0^8 + 6 b_0^4 b_1^6 F + b_1^{12} F^2}{4 b_0^6 b_1^2}, \frac{-b_0^{12} + 15 b_0^8 b_1^6 F + 9 b_0^4 b_1^{12} F^2 + b_1^{18} F^3}{8 b_0^9 b_1^3} \right),
\ee
and the location for $\boxminus(Q \boxplus Q)$ is $(x_2, -y_2)$. Hence, these another global sections can support the point which parameterizes another $U(1)$ cover. 

Let us construct a spectral surface which describes a vector bundle with a structure group $S(U(1) \times U(1) \times U(3))$. For the parameterization of the $U(3)$ factor, we can make use of the parameterization of \eqref{14-0}--\eqref{14-5} for the case of $S(U(1) \times U(3)) \subset SU(4)$ vector bundle. If the spectral surface equation for the $SU(4)$ is parameterized by $a^{\prime}_{r}, (r=0, 2, 3, 4)$, then the parameterization for the $S(U(1) \times U(3))$ spectral cover becomes
\bea
a_0^{\prime} &=& b_0^2 d,\\
a_2^{\prime} &=& b_0 d_2 - b_1^2 d,\\
a_3^{\prime} &=& b_0 d_3 + b_1 d_2,\\
a_4^{\prime} &=& b_1 d_3.
\eea
with the elliptic fiber equation \eqref{Wei} with $f$ and $g$ tuned to $b_1^2 F$ and $-b_0^2 F$ respectively. The equation embedded into $Sp(4)$ becomes
\be
(b_0^2 - b_1^2 x)(-b_0^2 d^2 - b_1^2 d_2^2 F - 2 b_0 b_1 d_2 d_3 F - b_0^2 d_3^2 F +( b_1^2 d^2  - 2 b_0 d d_2 )x - (d_2^2 + 2 b_1 d d_3)x^2 +d _3^2 x^3) = 0 \label{U(3)}
\ee
The second factor \eqref{U(3)} describes the $U(3)$ part. Then, the group sum of the three points described by the spectral surface of the $U(3)$ factor can be $Q$ or $\boxminus Q$. In order to make the group sum of the five points from the spectral cover of $S(U(1) \times U(1) \times U(3))$ the zero section, the group sum of the two points for the two $U(1)$ factor should be $\boxminus Q$ or $Q$. This can be achieved if one chooses the two points as $Q$ and $\boxminus (Q \boxplus Q)$, or $\boxminus Q$ and $Q \boxplus Q$. However, there are no difference between the two cases if one only look at the equations embedded into $Sp(5)$. Then, the equation which is embedded in the $Sp(5)$ spectral cover becomes
\bea
&&(b_1^2 x - b_0^2) (4b_0^6 b_1^2 x - (b_0^8 + 6 b_0^4 b_1^6 F + b_1^{12} F^2))(-b_0^2 d^2 - b_1^2 d_2^2 F \nn \\
&& - 2 b_0 b_1 d_2 d_3 F - b_0^2 d_3^2 F +( b_1^2 d^2  - 2 b_0 d d_2 )x - (d_2^2 + 2 b_1 d d_3)x^2 +d _3^2 x^3) = 0 \label{1x1x3}
\eea
The first two factors of \eqref{1x1x3} describe the two $U(1)$ factors. The location of $Q$ is completely determined if one fixes $b_0/b_1$, but this fixing is not enough for fixing the point from the second factor of \eqref{1x1x3}. Hence, the two points parameterizing the two $U(1)$ covers are independent.  

The next task is to find a solution for $a_r, (r=0,2,3,4,5)$ of the $SU(5)$ spectral cover equation \eqref{su5cover} which describes \eqref{1x1x3}. Indeed, one can find a unique solution 
\bea
a_0 &=&-b_0^2 (b_0^4 d + b_1^6 d F - 2 b_0 b_1^4 d_2 F - 2 b_0^2 b_1^3 d_3 F),\label{113-0}\\
a_2 &=& 3 b_0^4 b_1^2 d - b_0^5 d_2 + b_1^8 d F - b_0 b_1^6 d_2 F,\\
a_3 &=& -2 b_0^3 b_1^3 d - b_0^4 b_1 d_2 - b_0^5 d_3 - b_1^7 d_2 F - b_0 b_1^6 d_3 F,\\
a_4 &=& 2 b_0^3 b_1^2 d_2 - b_0^4 b_1 d_3 - b_1^7 d_3 F,\label{113-5}\\
a_5 &=& 2 b_0^3 b_1^2 d_3.
\eea
Therefore, the spectral surface equation \eqref{su5cover} with the parameterization \eqref{113-0}--\eqref{113-5} describes a spectral surface for the $S(U(1) \times U(1) \times U(3))$ vector bundle.


\section{F-theory interpretation}
 
In Section \ref{sec:spectral}, we have constructed spectral covers for vector bundles whose structure groups are $S(U(1) \times U(4))$, $S(U(2) \times U(3))$ and $S(U(1) \times U(1) \times U(3))$ in heterotic string theory. Then, 
an unbroken $U(1)$ symmetry is expected to be present in a four- dimensional low energy effective field theory. Although the presence of the unbroken $U(1)$ symmetry is implicit in heterotic string compactifications, its presence will be more explicit in F-theory compactifications. In this section, we will 
explicitly see evidences for the presence of the additional $U(1)$ symmetry in a four-dimensional low energy effective theory from F-theory compactifications.  


\subsection{Dual F-theory picture}

We first review the setup of F-theory compactifications which are dual to the heterotic string compactifications discussed in Section \ref{sec:spectral}. A prototype of heterotic -- F-theory duality is a eight-dimensional duality. Heterotic string on a torus is dual to F-theory on an elliptically fibered K3 surface \cite{Vafa:1996xn}. Then, one can apply the adiabatic argument \cite{Vafa:1995gm} to consider lower-dimensional dualities. In Section \ref{sec:spectral}, we have considered $\mathcal{N}=1$ supersymmetric heterotic string compactifications on Calabi--Yau threefolds $Z_3$ which have an elliptic fiber over a surface $S$. 
By the adiabatic argument, the same four-dimensional effective field theories can also be realized by F-theory compactifications on Calabi--Yau fourfolds $X_4$ which have a K3 fiber over the same surface $S$. 
The fibered K3 surface itself should have an elliptic fiber to hold the duality, and hence the Calabi--Yau fourfolds $X_4$ also have an elliptic fiber over threefolds $B_3$. The precise dictionary on the four-dimensional duality has been discussed extensively in \cite{Friedman:1997yq, Bershadsky:1997zs}.

In general, the elliptic fiber degenerates at some loci on $B_3$. The degeneration of the elliptic fiber indicates the presence of $[p, q]$ 7-branes at the degeneration locus. One can geometrically engineer non-Abelian gauge symmetries on 7-branes by requiring certain types of the singularities \cite{Morrison:1996na, Morrison:1996pp, Bershadsky:1996nh, Katz:2011qp}. The non-Abelian gauge symmetries on 7-branes correspond to the unbroken non-Abelian gauge symmetries in the dual heterotic string theory. The separation of the 7-branes can break the non-Abelian gauge symmetries. This can be achieved by taking certain complex structure moduli of $X_4$. One the other hand, the breaking of the non-Abelian gauge symmetries in heterotic string theory can be achieved by turning on a vector bundle. Therefore, the bundle moduli associated to the vector bundle in heterotic string theory correspond to the complex structure moduli of $X_4$ in F-theory. 

In Section \ref{sec:spectral}, we have discussed the decomposition of the spectral covers in heterotic string theory, assuming that the supergravity limit is valid. This is because the analysis is based on the dimensional reduction. In order to make the classical geometry valid, the area of the torus should be sufficiently large compared to the string scale $\alpha^{\prime}$. This case can be captured by the stable degeneration limit of the fibered K3 surface in F-theory compactifications \cite{Morrison:1996pp, Friedman:1997yq}. In this limit, the K3 surface splits into two rational elliptic surfaces\footnote{In some literatures, a rational elliptic surface is called as a del Pezzo nine surface.} $W_1$ and $W_2$ which share a common elliptic curve $E$, namely ${\rm K3} = W_1 \cup_E W_2$. The elliptic curve can be identified with the elliptic curve in heterotic string theory. After fibering over the surface $S$, the Calabi--Yau fourfolds $X_4$ can be expressed as
\be 
X_4 = Y_1 \cup_{Z_3} Y_2 \label{stabledeg}.
\ee
where $Y_i$ denotes a 
$W_i$ fiber over the base $S$ for each $i=1,2$. Each rational elliptic surface may describe each $E_8$ part of the $E_8 \times E_8$ gauge theory in heterotic string theory. The separation of two $E_8$ is achieved by the geometrical splitting in the stable degeneration limit. Since we have concentrated on one $E_8$, 
we will focus only on $Y_1$. 

One of the nice features of F-theory compactifications is that both the bundle moduli and the complex structure moduli of Calabi--Yau threefolds $Z_3$ in heterotic string theory can be captured by the complex structure moduli of $Y_1$ in \eqref{stabledeg}. In order to decompose the spectral surface in heterotic string theory, we have tuned both the bundle moduli and the complex structure moduli. In F-theory picture, this tuning will map to the tuning of the complex structure moduli only. The explicit map between the bundle moduli of a vector bundle and the complex structure moduli of $Y_1$ has been already known in \cite{Katz:1997eq, Berglund:1998ej}. Let us see how the spectral data $a_r$ of the spectral surface \eqref{su5cover} appear in the defining equation of $Y_1$. In a stable degeneration limit, the defining equation of $Y_1$ which has an $A_4$ singularity is
\be
y^2 = x^3 + f x z^4 + g z^6 + (a_0 z^5 + a_2 z^3 x + a_3 z^2 y + a_4 z x^2 + a_5 x y),
\label{defeq}
\ee
where $z$ is an affine coordinate of the base $\P^1$ in the rational elliptic surface $W_1$. $a_r, (r=0,\cdots, 5)$ and $f, g$ are global holomorphic sections over the base $S$,
\be
a_r \in \Gamma(S; \cO_S (r K_S + \eta)), \quad f \in \Gamma(S; \cO_S (-4K_S)), \quad g \in \Gamma(S; \cO_S( -6 K_S) ),
\ee
where a divisor $\eta$ is related to a normal bundle \cite{Rajesh:1998ik}
\be
c_1(N_{S|B_3}) = 6 K_S + \eta.
\ee
Indeed, the complex structure moduli $a_r$ appeared in \eqref{defeq} precisely map to the spectral data $a_r$ of the spectral surface \eqref{su5cover} in heterotic string theory. Furthermore, $f$ and $g$ in \eqref{defeq} map to the complex structure moduli $f, g $ in \eqref{Wei} of $Z_3$ respectively. Therefore, one can explicitly check whether F-theory compactifications on \eqref{defeq} will have an additional $U(1)$ symmetry by inserting the parametrization we have found in Section \ref{sec:spectral}.

We move on the condition for the number of Abelian vector multiplet in four-dimensional effective field theories from F-theory compactifications. 
We focus on vector fields which correspond to $U(1)$ symmetries inside $E_8 \times E_8$ gauge symmetries in the dual heterotic string theory. This type of vector fields comes from the dimensional reduction of the three-form $C_3$ in the dual M-theory compactifications\footnote{There is another type of vector fields associated to $U(1)$ symmetries from F-theory compactifications. The other type comes from the dimensional reduction of the RR four-form $C_4$ in Type IIB string theory. Hence, the number of the Abelian vector multiplets whose vector fields originate from $C_4$ is $h^{2,1}(B_3)$. The vector fields propagates in the ``bulk" $B_3$.}. However, not all the topological two-forms in $H^{2}(X_4; \Z)$ correspond to the Abelian vector multiplets \cite{Morrison:1996pp, Dasgupta:1999ss}. Since $X_4$ has an elliptic fiber over the base $B_3$, $\pi_{X_4}^{\prime}: X_4 \rightarrow B_3$, the topological two-forms $\omega$ which generates the vector fields in the Abelian vector multiplets are in $H^{2}(X_4; \Z)$ but not in $H^{2}(B_3; \Z)$ nor in $H^{0}(B_3; R^{2}\pi_{X_4}^{\prime}{}_{\ast}\Z)$. Hence, the number of Abelian vector multiplet whose vector field originate from the three-form $C_3$ is\footnote{When $X_4$ has singularities which support non-Abelian gauge symmetries, the definition of $h^{1,1}(X_4)$ becomes subtle. In that case, one may consider the $(1,1)$-type Hodge number of a resolved Calabi--Yau fourfold $\tilde{X}_4$. Such resolutions can be performed by using toric methods as in \cite{toric1, bottomup, toric2}, or by direct resolution from a defining equation as recently shown in \cite{resolution}. Then, the rank of the non-Abelian gauge symmetries can be understood as the number of the blow up divisors for the resolution of the singularities, which contribute to $h^{1,1}(\tilde{X_4})$.} 
\be
n_{U(1)} = h^{1,1}(X_4) - h^{1,1}(B_3) -1. \label{uone} 
\ee
In the case where $X_4$ has a K3 fiber over the base $S$, $\pi_{X_4}: X_4 \rightarrow S$, the two-forms $\omega$ can be studied by $H^{0}(S; R^{2}\pi_{X_4}{}_{\ast}\Z)$ \cite{Curio:1998bva, Donagi:2008ca, Hayashi:2008ba}.  
Then, global sections of the local system $R^2\pi_{X_4 \ast}\Z$ correspond to the monodromy invariant two-cycles in the fiber $\pi_{X_4}$. In the stable degeneration limit, the K3 surface splits into two rational elliptic surfaces as in \eqref{stabledeg}. Hence, the existence of $\omega$ can be rephrased as the existence of monodromy invariant two-cycles in $W_1$. Therefore, the correspondence between $U(1)$ symmetries and the geometry of F-theory compactifications is 
\be
\text{$U(1)$ symmetries on 7-branes} \leftrightarrow \text{monodromy invariant two-cycle in $W_1$} . \label{uone1}
\ee
This type of vector fields correspond to the gauge fields on 7-branes. 
We will analyze the monodromy invariant two-cycles in $W_1$ to see the existence of $U(1)$ symmetries. 


\subsection{$E_6$ gauge theory without decomposition}
\label{sec:E6-0}

We have seen that the existence of an $U(1)$ symmetry is equivalent to the existence of a monodromy invariant two-cycle in $W_1$ in F-theory compactifications. 
The monodromy of two-cycles in $W_1$ can be explicitly calculated from the defining equation of $Y_1$ \eqref{defeq}. Then, we can check the existence of a monodromy invariant two-cycle if one applies the parameterizations of the $(1+4)$, $(2+3)$ or $(1+1+3)$ decomposition to \eqref{defeq}. In the explicit calculation of the monodromy, we only consider an $E_6$ gauge theory with $(1+2)$ decomposition for simplicity. The behavior of the monodromy locus for $SU(5)$ gauge theories will be studied in Section \ref{sec:su5}. The study of the monodromy of the $E_6$ gauge theory will support the observation from the results of the monodromy locus in the $SU(5)$ gauge theory. First, we obtain the monodromy of two-cycles without taking the decomposition limit.

An explicit computation of monodromy invariant two-cycles in the K3 fiber has been carried out in \cite{Hayashi:2010zp}. The monodromy of two-cycles was derived from the motion of 7-branes associated to loops in the base $S$. 
The application of the method to the rational elliptic surface fibration case has been also done in \cite{unpublished}. The application is straightforward and we follow the same notation and procedure in \cite{Hayashi:2010zp} in this paper.

\begin{figure}[t]
\begin{center}
\begin{tabular}{c}
\includegraphics[scale=0.35]{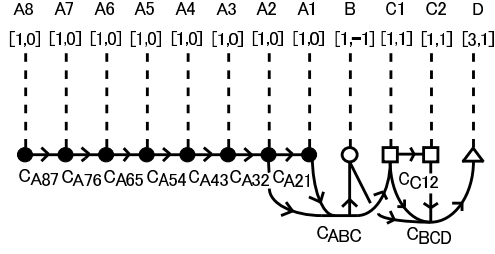} 
\end{tabular}
\caption{The configuration of the twelve 7-branes and two-cycles in the rational elliptic surface $W_1 = \pi_{Y_4}^{-1}(b)$.}
\label{fig:2-cycles}
\end{center}
\end{figure}

 We first set the notation of two-cycles in a rational elliptic surface. 
If one focuses on a point in the base $S$, we have in total twenty four 7-branes \emph{locally} in the case of the K3 fibration. Locally because if we consider the entire $S$, the global structure would reveal that they are not all independent branes. The charges of the 7-branes and the string junction or two-cycle configuration between them have been determined in \cite{DeWolfe:1998pr}. In the stable degeneration limit, the twenty four 7-branes are separated into two groups. Each of $W_1$ and $W_2$ has twelve 7-branes. 
We assign names $A8, \cdots, A1, B, C1, C2, D$ to individual twelve $[p, q]$ 7-branes at a base point $b \in S$.  
We also assign names $C_{A87}, C_{A76}, \cdots$, $C_{ABC}$, $C_{C12}$, $C_{BCD}$ to string junctions between 7-branes and corresponding two-cycles in the rational elliptic surface $W_1 = \pi_{Y_4}^{-1}(b)$. The information is summarized in Figure \ref{fig:2-cycles}. 
%

\begin{figure}[t]
\begin{center}
\begin{tabular}{c}
\includegraphics[scale=0.35]{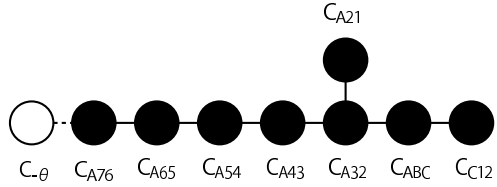} 
\end{tabular}
\caption{The extended Dynkin diagram of the $E_8$ Lie algebra. Each node corresponds to a two-cycle in Figure \ref{fig:2-cycles}. }
\label{fig:dynkin}
\end{center}
\end{figure}

Actually, not all of the ten two-cycles in Figure \ref{fig:2-cycles} 
are independent in $W_1$. Note that a rational elliptic surface has ten two-cycles. However, two of them are an elliptic fiber and a base $\P^1$. Surely, none of the two-cycles in Figure \ref{fig:2-cycles} correspond to the elliptic fiber or the base $\P^1$. 
Namely, only eight out of the ten two-cycles 
are independent, so we choose $C_{A76}, \cdots, C_{ABC}, C_{C12}$ as independent eight two-cycles. In fact, these eight two-cycles 
correspond to the simple roots of the $E_8$ Lie algebra. One can also explicitly confirm that the intersection form of the eight two-cycles is the negative of the Cartan matrix of $E_8$ Lie algebra. Furthermore, a two-cycle $C_{-\theta}$ corresponding to a minimal root is constructed by a linear combination of the simple roots
\be
C_{-\theta} = -(2 C_{A76} + 3 C_{A65} + 4 C_{A54} + 5C_{A43} + 6 C_{A32} + 4 C_{ABC} + 3 C_{A21} + 2 C_{C12}).
\ee
Then, the intersection structure between two-cycles $C_{A76}, \cdots, C_{ABC}, C_{C12}$ and $C_{-\theta}$ gives the extended Dynkin diagram for the $E_8$ Lie algebra as in Figure \ref{fig:dynkin}. 
On the other hand, the other two two-cycles $C_{A87}$ and $C_{BCD}$ can be written by a linear combination of $C_{-\theta}$ and a boundary of three-dimensional cells\footnote{For the cell decomposition, see for example \cite{BT}.}. In fact, those eight two-cycles are the ones which are related to $U(1)$ symmetries in F-theory compactifications. Since the elliptic fiber and the base $\P^1$ are related to $H^{0}(B_3; R^2\pi^{\prime}_{X_4}{}_{\ast}\Z)$ and $H^{2}(B_3; \Z)$ respectively, they do not contribute to \eqref{uone}. Hence, the relevant two-cycles for the monodromy are $C_{A76}, \cdots, C_{ABC}, C_{C12}$ and $C_{-\theta}$. 

$$* * *$$

We move on to a more specific model, namely an $E_6$ gauge theory on 7-branes. An $E_6$ gauge symmetry can be realized if eight 7-branes $A5, \cdots, A1, B, C1, C2$ coincide with each other \cite{Johansen:1996am, Gaberdiel:1997ud}. Therefore, there are six vanishing two-cycles $C_{A54}, C_{A43}, C_{A32}, C_{A21}, C_{ABC}$ and $C_{C12}$. The other two-cycles have finite size. Hence, we will consider the monodromy of the two-cycles, $C_{A65}, C_{A76}$ and also $C_{-\theta}$.  
We may expect that  monodromy group would generically include $S_3$, the Weyl group of a structure group $SU(3)_{\perp}$ in the case of the breaking $E_8 \supset E_6 \times \langle SU(3)_{\perp} \rangle$. We will see that the monodromy coming from 
branch points in 
a subspace indeed yields generators of $S_3$.

The $E_6$ singularity can be realized by the following geometry \cite{Bershadsky:1996nh}. 
Note that eq.~\eqref{defeq} represents $Y_1$ which has an $A_4$ singularity. Then, one can engineer an $E_6$ singularity by setting $a_5=a_4=0$ in \eqref{defeq}, namely
\be
y^2 = x^3 + f x z^4 + g z^6 + (a_0 z^5 + a_2 z^3 x + a_3 z^2 y),
\label{defeq1}
\ee
The eq.~\eqref{defeq1} has an $E_6$ singularity along $z=0$. 
The locations of 7-branes are characterized by the degeneration of the elliptic fiber. The loci can be captured by the discriminant of \eqref{defeq1}. After making the defining equation \eqref{defeq1} into the Weierstrass form by the coordinate changes for $y$, one can compute the discriminant of \eqref{defeq1} and the result is
\be
\Delta = z^8 \Delta^{\prime}, \label{dis1}
\ee
where $\Delta^{\prime}$ is
\bea
\Delta^{\prime} &=& \frac{27}{16}a_3^4 + \left(\frac{27}{2}a_0a_3^2 + 4a_2^3\right)z + \left(\frac{27}{2} a_3^2 g + 27 a_0^2 + 12a_2^2 f\right)z^2 \nn \\
&&+ \left( 54 a_0 g + 12 a_2 f^2\right) z^3 + (4f^3+ 27 g^2)z^4. \label{dis2}
\eea
Note that the $z^8$ factor in \eqref{dis1} describes the eight 7-branes $A5, \cdots, A1, B, C1, C2$ which realize the $E_6$ gauge symmetry at the origin. 
The motion of the other four 7-branes $A8, A7, A6, D$ is governed by the discriminant \eqref{dis2}. 

When one simply sets $f$ and $g$ to be zero, then, the discriminant \eqref{dis1} reduces to a degree ten polynomial in $z$. Hence, only the 7-branes 
$A7, \cdots, A1, B, C1, C2$ of $E_8$ takes part in the calculation in that case. This is what Higgs bundle description captures, and is related to 8D gauge theory region \cite{Hayashi:2010zp}. However, we indeed have two other 7-branes $A8, D$ and cannot neglect their effects generically.  
 
The locations of the four 7-branes change according to \eqref{dis2} as one moves along a path in the base $S$. 
In particular, when one encircles a branch locus in the base $S$, the locations of 
7-branes are interchanged with each other and it may cause monodromy of two-cycles. The branch locus can be studied by the discriminant of \eqref{dis2}
\be
\tilde{\Delta}^{\prime} = -19683 A^3 B, \label{disofdis}
\ee
where $A$ and $B$ are given by
\bea
A &=& 4 a_0 a_2 f - a_3^2 f^2 - 4 a_2^2 g,\\
B &=& 4a_0^3a_2^3 + 27 a_0^4 a_3^2 - 4a_2^6 g - 36 a_0 a_2^3 a_3^2 g - 54 a_0^2 a_3^4 g + 27 a_3^6 g^2 \nn \\
&&+ 4 a_0 a_2^5 f + 30 a_0^2 a_2^2 a_3^2 f + 18 a_2^2 a_3^4 g f - a_2^4 a_3^2 f^2 - 24 a_0 a_2 a_3^4 f + 4a_3^6 f^3, \label{B}
\eea
We call the locus $\{ \tilde{\Delta}^{\prime} = 0 \}$ as a monodromy locus.

Although we consider loops in the base $S$, 
it turns out to be more convenient to use loops in the moduli space $M \equiv (a_0, a_2, a_3, f, g)$. There is one-to-one map between loops in $S$ and the ones in $M$. Since $(a_0, a_2, a_3, f, g)$ are generic sections ({\it i.e.} the complex parameters in the sections are not tuned), a $2\pi$ rotation with a very small radius which encircles branch points in the base $S$ always maps to a $2\pi$ rotation which encircles branch points in the moduli space $M$. Hence, one can consider $(a_0, a_2, a_3, f, g)$ as complex parameters effectively for the computation of the monodromy. Then, the parameterization of $(a_0, a_2, a_3, f, g)$ are actually redundant because of the two rescaling for $(x, y, z)$ in \eqref{defeq1}. Instead of working on gauge invariant objects, we just fix two of $(a_0, a_2, a_3, f, g)$ by using the redundancy. In the later analysis, we fix the ``gauge" for $(a_3, f)$, so that the remaining physical degrees of freedom are $M_f \equiv (a_0, a_2, g)$.

In the analysis of the monodromy, we will fix some parameters for the simplicity of the calculation. The most convenient choice is going to 8D gauge theory region \cite{Hayashi:2010zp}, in which all the parameters in $M$ should be appropriately parameterized to be consistent with the scaling. Using ``gauge" degree of freedom, we fix $(a_3, f) = (i \epsilon_K^3 \delta, -1)$ where $\epsilon_K$ and $\delta$ represent small real numbers. And to copy with the scale we appropriately fix two parameters in $M_{f}$ for the simplicity. 
The further scaling by $\delta$ makes the $A6$ 7-brane locate closely at $z=0$, which simplifies the calculations. 
Certainly, the analysis of the whole moduli space of $(a_0, a_2, g)$ is necessary to ensure the whole monodromy group. 
However, Eq.~\eqref{disofdis} itself does factorize when one adopts the decomposition conditions of all the cases in Section \ref{sec:spectral}, and the factors always include a squared factor which would be important for the reduction of the monodromy. Hence, we may expect that the result we will find may not change if one works in the full moduli space $(a_0, a_2, g)$. 

\begin{figure}[t]
\begin{center}
\begin{tabular}{cc}
  \includegraphics[scale=0.25]{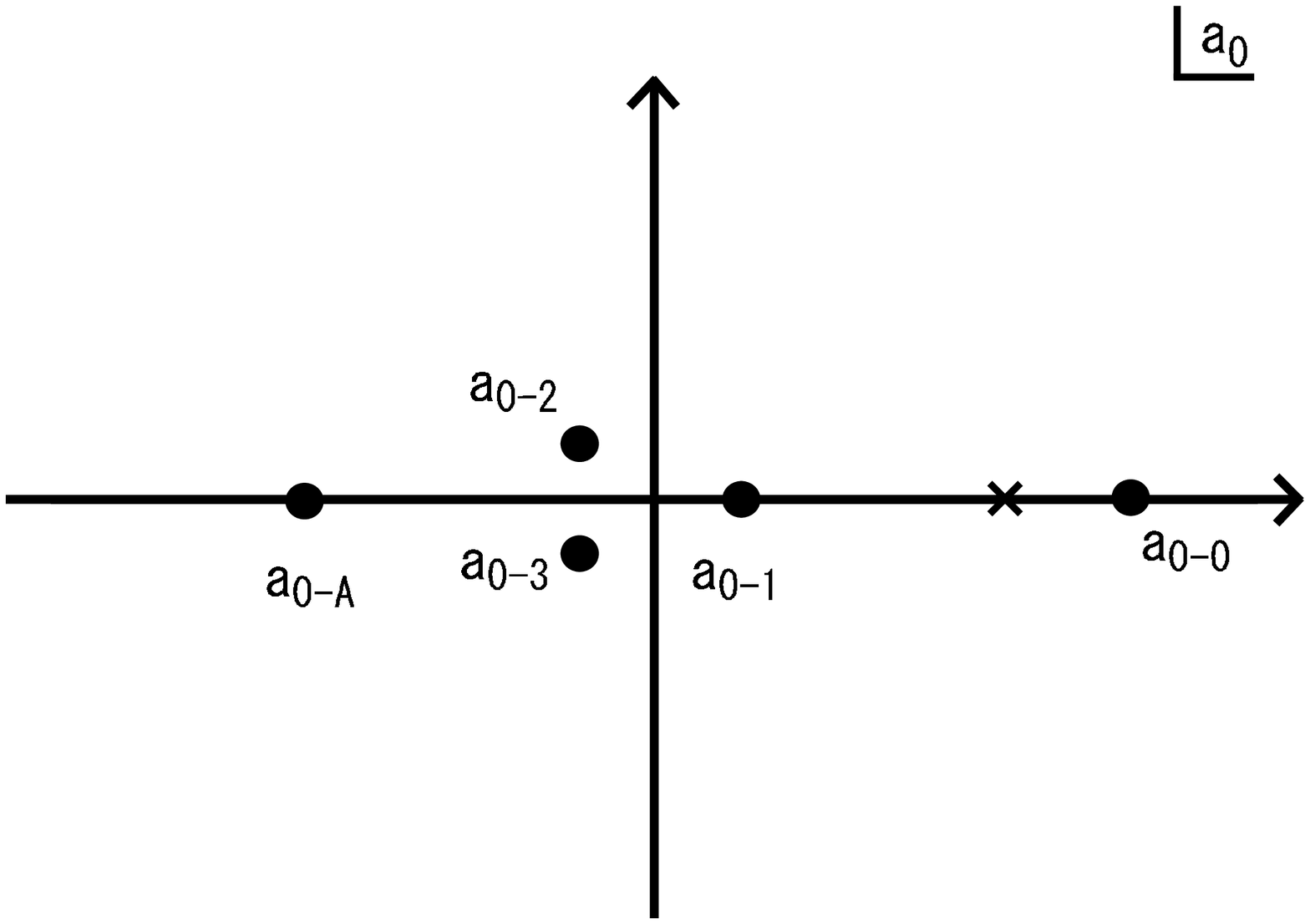} &  
  \includegraphics[scale=0.25]{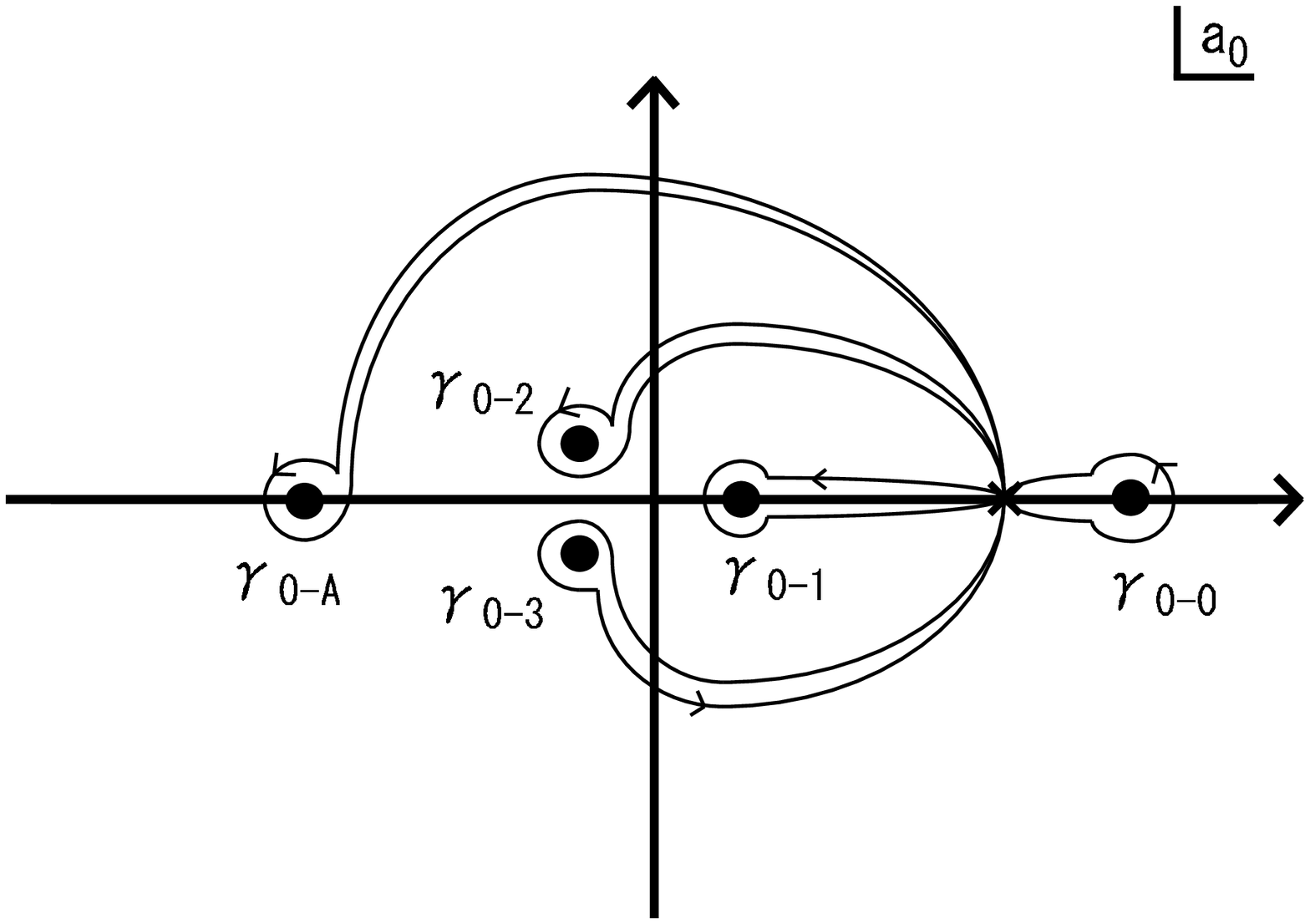} \\
 (a) & (b)
\end{tabular}
\caption{The left figure (a) shows the five branch points in the $a_0$-plane. The cross mark represents the base point $b$. The right figure (b) shows the loops corresponding to the five branch points in the $a_0$-plane.}
\label{fig:a0}
\end{center}
\end{figure}

As for the analysis without any tuning, we will consider an $a_0$-plane where $a_2$ and $g$ are fixed to
\be
(a_2, g) = (\epsilon_K^2, 1), \quad \epsilon_K \in \R, \quad \epsilon_K \ll 1. \label{a0-plane}
\ee
Then, the monodromy locus \eqref{disofdis} is a product of cubic of a degree one polynomial $A$ of $a_0$ and a degree four polynomial $B$ of $a_0$. The three roots from $A^3=0$ share the same point in the $a_0$-plane, and hence there are five branch points in the $a_0$-plane in total. The five branch points are depicted in Figure \ref{fig:a0} (a). $a_{0-A}$ represents a triple point from the $A^3=0$ factor in \eqref{disofdis}. The other four branch points $a_{0-0}, a_{0-1}, a_{0-2}, a_{0-3}$ are single points from the $B=0$ factor in \eqref{disofdis}. 

\begin{figure}[t]
\begin{center}
\begin{tabular}{c}
\includegraphics[scale=0.3]{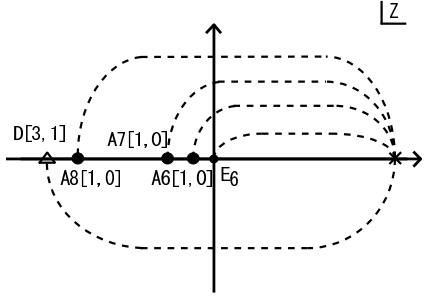}
\end{tabular}
\caption{The configuration of 7-branes at the base point $b$.}
\label{fig:base_pt}
\end{center}
\end{figure}

As with the computation in \cite{Hayashi:2010zp}, we choose a base point $b$ as $a_0 = 1$ in the $a_0$-plane \eqref{a0-plane}. 
At the base point $b$, the charge of the four $[p,q]$ 7-branes away from the origin are precisely the ones written in Figure \ref{fig:2-cycles}. The explicit location of the four 7-branes $A8, A7, A6, D$ in the $z$-plane and the choice of the branch cuts at the base point $b$ is depicted in Figure \ref{fig:base_pt}. 
From the base point $b$, we consider five loops which encircle each five branch points. We label them as $\gamma_{0-A}, \gamma_{0-0}, \gamma_{0-1}, \gamma_{0-2}, \gamma_{0-3}$ in Figure \ref{fig:a0} (b).

We only present the result of the computation of the monodromy associated to the five loops. The method of the calculation is explained in \cite{Hayashi:2010zp} in detail. 
From the direct computation, the monodromy associated with the loop $\gamma_{0-A}$ gives trivial monodromy \cite{Hayashi:2010zp}. Therefore, we do not need to consider the $A=0$ component of the monodromy locus anymore. If a loop $\gamma_{A}$ is homotopic to a loop of the form
\be
\gamma_{A} = \gamma_{0}^{-1} \circ \gamma_{0-A} \circ \gamma_{0},
\ee
in the moduli space $(a_0, a_2, g)$ except for $A=0$ and $B=0$ for any loop $\gamma_{0}$, then the monodromy associated to the loop $\gamma_{A}$ is trivial. This is because the monodromy associated to the loop $\gamma_{0-A}$ is trivial. Therefore, the monodromy locus $A=0$ have no effect on the other monodromy. 

Hence, 
let us consider the monodromy from the other loops $\gamma_{0-0}, \gamma_{0-1}, \gamma_{0-2}, \gamma_{0-3}$ in Figure \ref{fig:a0} (b). The results have been already obtained in \cite{unpublished}. From the explicit calculation, one can show that the monodromy associated to $\gamma_{0-0}$ is 
\bea
\tilde{C}_{A65} &=& C_{A65} + C_{A76},\label{0-1}\\
\tilde{C}_{A76} &=& -C_{A76},\label{0-2}\\
\tilde{C}_{-\theta} &=& C_{-\theta} + C_{A76}, \label{0-3}
\eea
where we put tilde for the two-cycles after one encircles a branch point. The transformations \eqref{0-1}--\eqref{0-3} are nothing but the Weyl reflection $W_{C_{A76}}$ with respect to the root $C_{A76}$. We express the corresponding monodromy element as $\rho(\gamma_{0-0}) = W_{C_{A76}}$. Similarly, the direct computation can show that the monodromy associated to all the loops $\gamma_{0-1}, \gamma_{0-2}, \gamma_{0-3}$ is the same and the explicit form is 
\bea
\tilde{C}_{A65} &=& C_{A65} ,\label{1-1}\\
\tilde{C}_{A76} &=& C_{A76} + C_{-\theta},\label{1-2}\\
\tilde{C}_{-\theta} &=& -C_{-\theta}. \label{1-3}
\eea
The transformations \eqref{1-1}--\eqref{1-3} are the Weyl reflection $W_{C_{-\theta}}$ with respect to the root $C_{-\theta}$. Therefore, two elements $W_{C_{A76}}, W_{C_{-\theta}}$ generate Weyl group $S_3$ as expected.


\subsection{$E_6$ gauge theory with $(1+2)$ decomposition - case I}
\label{sec:case1}

We next turn to the decomposition in Section \ref{sec:Hetcase1}. We apply the parametrization \eqref{a3factor} to the parameters $a_0, g$ in \eqref{defeq1}. The simple ansatz \eqref{a3factor} becomes\footnote{Similar geometry is discussed in \cite{Braun:2011zm} in order to construct a four-cycle which is Poincar\'e dual to $G$-flux in F-theory compactifications. In \cite{Braun:2011zm}, they required that $a_0$ is factorized for the construction of a four-cycle where we can turn on the G-flux. In the parameterization \eqref{case1} (and also \eqref{case2-1} in Section \ref{sec:case2}), the factorization of $a_0$ also occurs for the presence of a monodromy invariant two-cycle. Hence, one may expect that the conditions \eqref{case1} (and also \eqref{case2-1}) not only realize a $U(1)$ symmetry in a low energy effective field theory but also generate a four-cycle for the G-flux which may induce the chiral spectrum \cite{progress}.}
\be
a_0 = a_{-3} a_3, \qquad g = a_{-3}^2, \label{case1}
\ee
and the other parameters $a_2, a_3, f$ are generic.  

In order to compute the monodromy of two-cycles under the parametrization \eqref{case1}, we insert the parameterization \eqref{case1} into \eqref{B}, and $B$ becomes, 
\be
B \rightarrow -(2 a_2 a_{-3} - a_3 f)^2 (a_2^4 + 8 a_2 a_3^3 a_{-3} - 4 a_3^4 f),\label{disofdis1}
\ee
Hence, a further factorization occurs in the case of the parametrization \eqref{case1}. We point out that the tuning of $g$ plays a non-trivial role for the factorization \eqref{B}. The tuning of $g$ will be never predicted if one only considers the spectral surface from the Higgs bundle since it is a subdominant parameter in the discriminant in 8D gauge theory region. 

\begin{figure}[t]
\begin{center}
\begin{tabular}{c}
\includegraphics[scale=0.3]{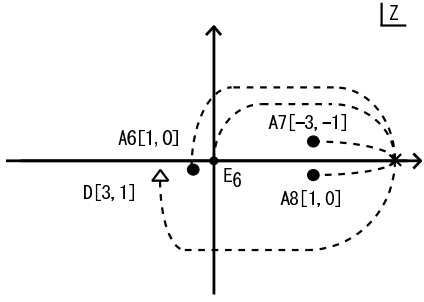}
\end{tabular}
\caption{The configuration of 7-branes at the base point $b_1$.}
\label{fig:b1}
\end{center}
\end{figure}

We only consider the monodromy in the $a_{-3}$-plane with $a_2$ fixed to $\epsilon_K^2$. $(a_3, f)$ can be ``gauge" fixed to $(i \epsilon_K^3 \delta, -1)$.  We choose the base point $b_1$ in the $a_{-3}$-plane as $a_{-3} = 1$. 
This $b_1$ is mapped to $a_0 = i \epsilon_K^3 \delta$ rather than $b$ in the $a_0$-plane. The configuration of the 7-branes at the base point $b_1$ is depicted in Figure \ref{fig:b1}. 
We can still make use of the basis two-cycles $C_{A76}, \cdots, C_{ABC}, C_{C12}$ for the computation of monodromy of loops starting from the base point $b_1$. In the $a_0$-plane, we can consider a path $\gamma^{\prime}$ from $a_0 = 1$ to $a_0 = i \epsilon_K^3 \delta$. Although the expressions (for examples, in terms of string jucntions) of the two-cycles $C_{A76}, \cdots, C_{ABC}, C_{C12}$ change 
according to the path $\gamma^{\prime}$, we can trace the two-cycles until the base point $b_1$. 
Then, we can use the same symbol for the basis two-cycles 
at the base point $b_1$. 
The configurations of the two-cycles $C_{A65}, \cdots, C_{A21}, C_{ABC}$ and $C_{C12}$ at $b_1$ are essentially the same as the ones in Figure \ref{fig:2-cycles}. The other two-cycles are summarized in Table \ref{tb:b1}. 
\begin{table}[t]
\begin{center}
\begin{tabular}{|c|c|c|c|c|c|c|c|c|}
\hline
& A6[1,0] & A5[1,0] & A4-1[1,0] & B[0, -1]& C1,2[2,1]& A7[-3,-1] & A8[1,0]& D[3,1]\\
\hline
$C_{A76}$ & -1 & 0 &  0 & 1 & 1 & 1 & 0 & 0\\
\hline
$C_{A87}$ & 0 & 0 &  0 & -1 & -1  & -1 & 1 & 0\\
\hline
$C_{BCD}$ & 0 & 0 &  0 & 1 & 1  & 0 & -1 & 1\\
\hline
\end{tabular}
\caption{The configuration of the two-cycles $C_{A76}, C_{A87}$ and $C_{BCD}$ at the base point $b_1$. Each number denotes the charges for the corresponding string junctions. The plus/minus sign associated to a string junction indicates that the string junction is stretched from/to a 7-brane. The number represents the number of the strings.}
\label{tb:b1}
\end{center}
\end{table}
Since the intersection form between the two-cycles does not change along the path $\gamma^{\prime}$, the eight two-cycles can still be considered as the simple roots of the $E_8$ Lie algebra. 

\begin{figure}[t]
\begin{center}
\begin{tabular}{cc}
  \includegraphics[scale=0.3]{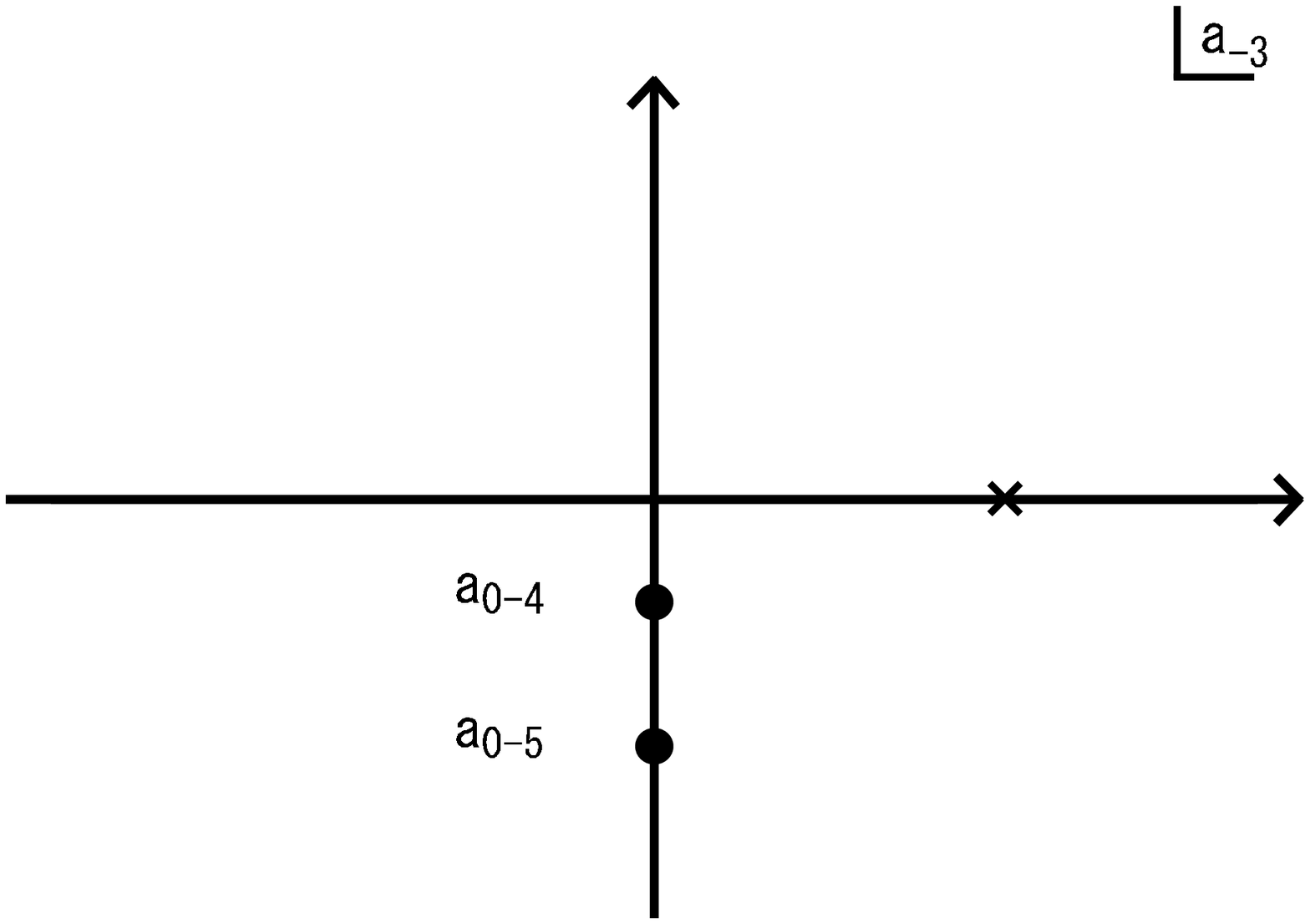} &  
  \includegraphics[scale=0.3]{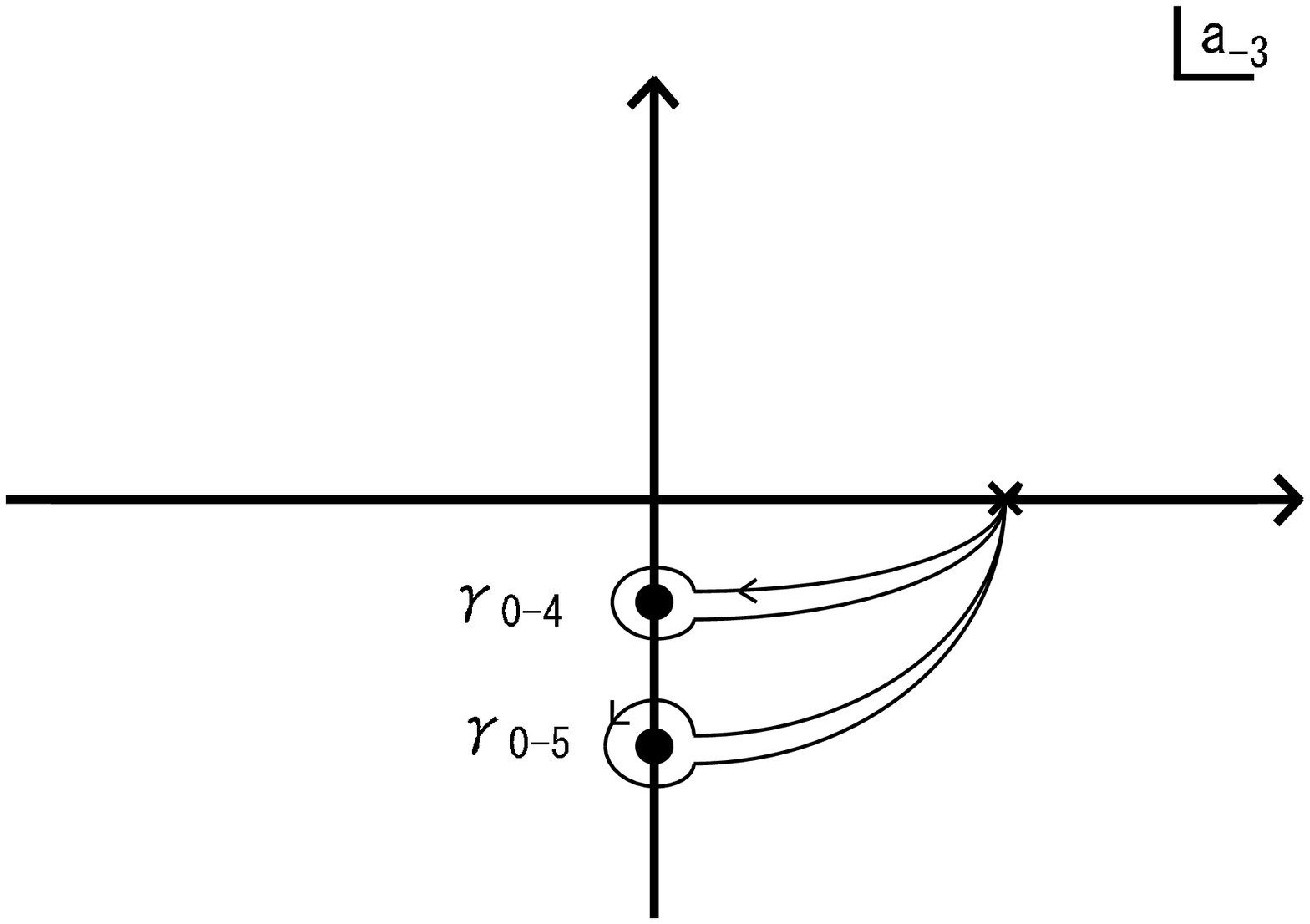} \\
 (a) & (b)
\end{tabular}
\caption{The left figure (a) shows the two branch points in the $a_{-3}$-plane. The cross mark represents the base point $b_1$. The right figure (b) shows the loops corresponding to the two branch points in the $a_{-3}$-plane.}
\label{fig:a-3}
\end{center}
\end{figure}

Let us compute the branch points from \eqref{disofdis1} in the $a_{-3}$-plane. We have two types of the branch points. One comes from the first factor of \eqref{disofdis1} and the two points coincide at one point. We express the branch point as $a_{0-4}$ in Figure \ref{fig:a-3} (a). 
The other branch point comes from the second factor of \eqref{disofdis1}. We express the branch point as $a_{0-5}$ in Figure \ref{fig:a-3} (b). Then, we consider loops which start from the base point $b_1$, encircle the branch points $a_{0-4}$ and $a_{0-5}$, and finally goes back to the base point $b_1$. They are labeled as $\gamma_{0-4}, \gamma_{0-5}$ respectively in Figure \ref{fig:a-3} (b). 

In the same way as in Section \ref{sec:E6-0}, one can explicitly compute the monodromy $\rho(\gamma_{0-4})$ and $\rho(\gamma_{0-5})$ associated to the loops $\gamma_{0-4}$ and $\gamma_{0-5}$. The results are
\bea
\rho(\gamma_{0-4}) &=& \mathbb{I}, \label{trivial1}\\
\rho(\gamma_{0-5}) &=& W_{C_{A76}},
\eea
where $\mathbb{I}$ denotes trivial monodromy. Therefore, the monodromy group from the branch points in $a_{-3}$-plane is generated by just one Weyl reflection $W_{C_{A76}}$. Hence, the monodromy in the $a_{-3}$-plane is $\Z_2$. At least, we can show that the monodromy is reduced from $S_3$ to $\Z_2$ in the $a_{-3}$-plane. 

Interestingly, the monodromy associated to a loop of the branch point form the squared factor in \eqref{disofdis1} gives the trivial monodromy \eqref{trivial1}. However, the triviality of the monodromy is not a consequence of a twice Weyl reflection. In fact, the mutually non-local A7, A8 and D 7-branes rotate together by more than $2\pi$ radian when one goes along the loop $\gamma_{0-4}$. After the non-trivial transformation of the two-cycles, the final expression of the two-cycles does not change at the base point $b_1$ and the monodromy associated to the loop $\gamma_{0-4}$ becomes trivial. 


We have observed the reduction of monodromy on F-theory side, at least in the $a_{-3}$-plane. This shows a power of string duality. Note that the parametrization \eqref{case1} is inferred from the analysis in heterotic string theory. It is highly non-trivial to find a parametrization which can reduce the monodromy only from the defining equation \eqref{defeq1}. However, if one applies the parametrization \eqref{case1} which is predicted by the spectral cover construction in heterotic string theory, the monodromy locus \eqref{disofdis} does factorize with squared factors and this would be important why the monodromy gets reduced.


\subsection{$E_6$ gauge theory with $(1+2)$ decomposition - case II}
\label{sec:case2}

Next, we apply the parametrization \eqref{14-0}--\eqref{14-5} and \eqref{holtuning} to the $E_6$ gauge theory in F-theory compactifications. The computation is basically the same as the one in Section \ref{sec:case1}. In the case of $E_6$ gauge theory, the parametrization of \eqref{14-0}--\eqref{14-5} and \eqref{holtuning} become
\bea
a_0 &=& d b_0^2, \quad
a_2 = -b_1^2 d + d_2 b_0, \quad
a_3 = d_2 b_1, \label{case2-1}\\
f &=& F b_1^2,\quad
g = - F b_0^2 \label{case2-2}
\eea
Inserting the parametrization \eqref{case2-1} and \eqref{case2-2} into the monodromy locus \eqref{B}, $B$ becomes
\be
B \rightarrow  (4d_2 d^3 b_0 - d^4 b_1^2 + 4d_2^4 F) (d_2 b_0^4 + 2d b_0^3 b_1^2 + d_2 b_1^6 F)^2. \label{disofdis3}
\ee
In this case, the monodromy locus $\{\tilde{\Delta}^{\prime}=0\}$ also factorizes in a non-trivial way. Note that the tuning \eqref{case2-2} of $f, g$ is crucial for the factorization in \eqref{disofdis3} also in this case. This factorization never occurs if one only considers the parametrization \eqref{case2-1} but not \eqref{case2-2}. 

\begin{figure}[t]
\begin{center}
\begin{tabular}{c}
\includegraphics[scale=0.3]{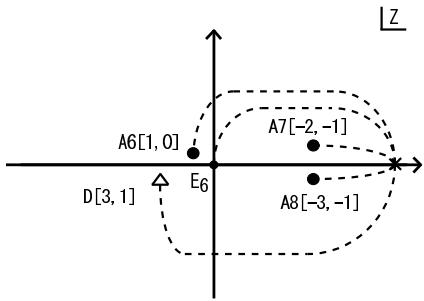}
\end{tabular}
\caption{The configuration of 7-branes at the base point $b_2$.}
\label{fig:b2}
\end{center}
\end{figure}

We then explicitly compute the monodromy from the monodromy locus  
\eqref{disofdis3} in the $d$-plane where $(d_2, b_0, b_1, F)$ fixed to $(\epsilon_K^2 \delta, i \epsilon_K, i \epsilon_K, \epsilon_K^{-2})$. This subspace is consistent with the ``gauge" fixing $(a_3, g) = (i \epsilon_K^3, -1)$ and the actual number of the parameters we fix is two. 
In the $d$-plane, we choose the base point $b_2$ as $d = 1$. The base point $b_2$ also does not map to the base point $b$ in the $a_0$-plane. The 7-brane configuration at the base point $b_2$ is depicted in Figure \ref{fig:b2}. 
In the same way as in Section \ref{sec:case1}, we can choose the basis two-cycles $C_{A76}, \cdots, C_{ABC}, C_{C12}$ as the ones obtained after the path $\gamma^{\prime \prime}$; $\gamma^{\prime \prime}: b \rightarrow b_2$ in the $a_0$-plane. The configurations of the two-cycles $C_{A65}, C_{A54}, C_{A32}, C_{A21}, C_{ABC}, C_{C12}$ are essentially the same as the ones in Figure \ref{fig:2-cycles}. The other two-cycles are summarized in Table \ref{tb:b2}.
\begin{table}[t]
\begin{tabular}{|c|c|c|c|c|c|c|c|c|}
\hline
& A6[1,0]& A5[1,0]& A4-1[1,0] & B[0,-1]& C1,2[2,1]& A7[-2,-1]& A8[-3,-1]& D[3,1]\\
\hline
$C_{A76}$ & -1 & 0 &  0 & 0 & 0  & 1 & -1 & 0\\
\hline
$C_{A87}$ & 0 & 0 &  0 & 1 & 1  & -1 & 2 & 0\\
\hline
$C_{BCD}$ & 0 & 0 &  0 & 1 & 1  & -1 & 1 & -1\\
\hline
\end{tabular}
\caption{The configuration of the two-cycles $C_{A76}, C_{A87}$ and $C_{BCD}$ at the base point $b_2$. Each number denotes the charges for the corresponding string junctions. The plus/minus sign associated to a string junction indicates that the string junction is stretched from/to a 7-brane. The number represents the number of the strings.}
\label{tb:b2}
\end{table}
Again, the eight two-cycles $C_{A76}, \cdots, C_{A21}, C_{ABC}, C_{C12}$ correspond to the simple roots of the $E_8$ Lie algebra. 

\begin{figure}[t]
\begin{center}
\begin{tabular}{cc}
  \includegraphics[scale=0.3]{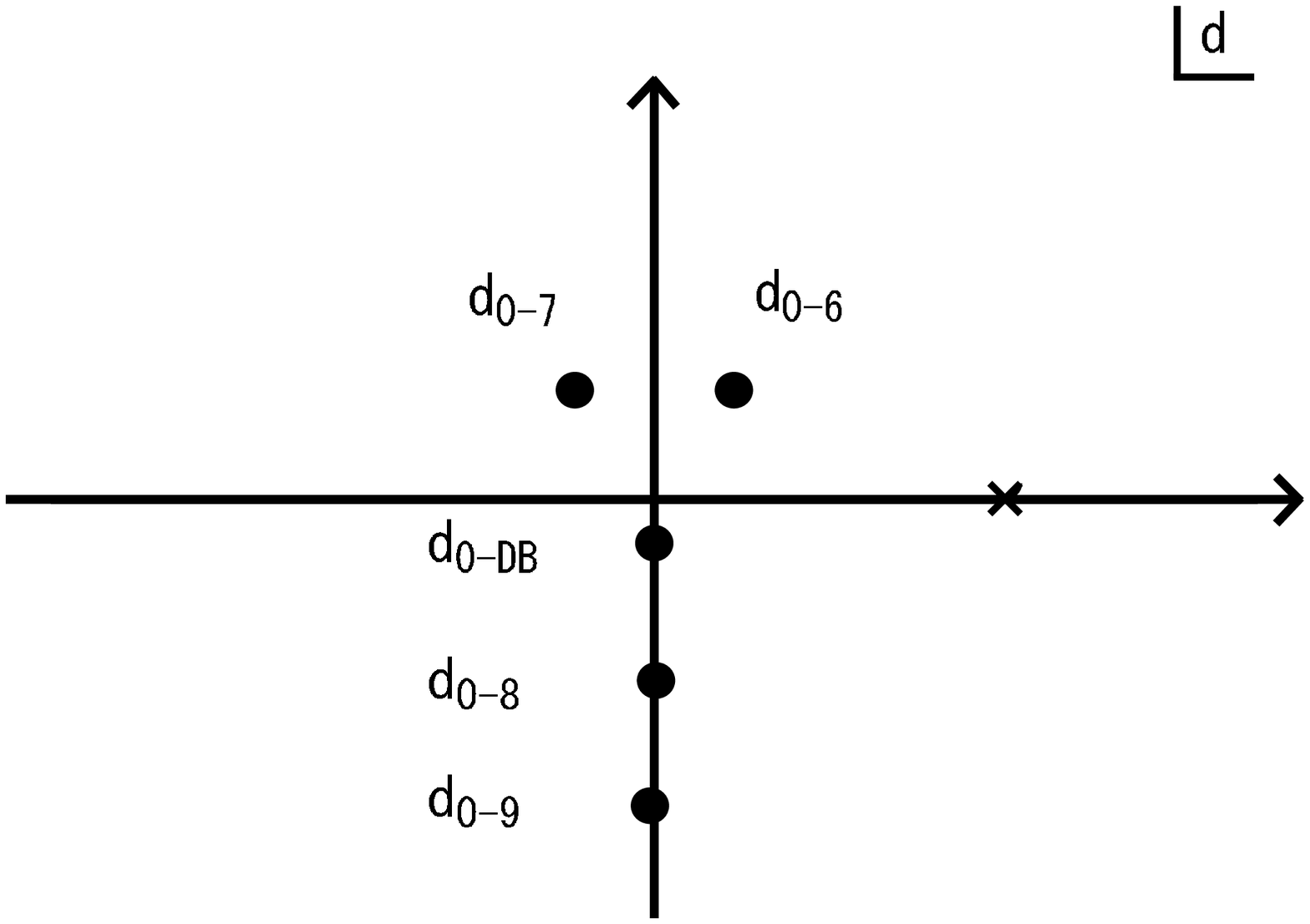} &  
  \includegraphics[scale=0.3]{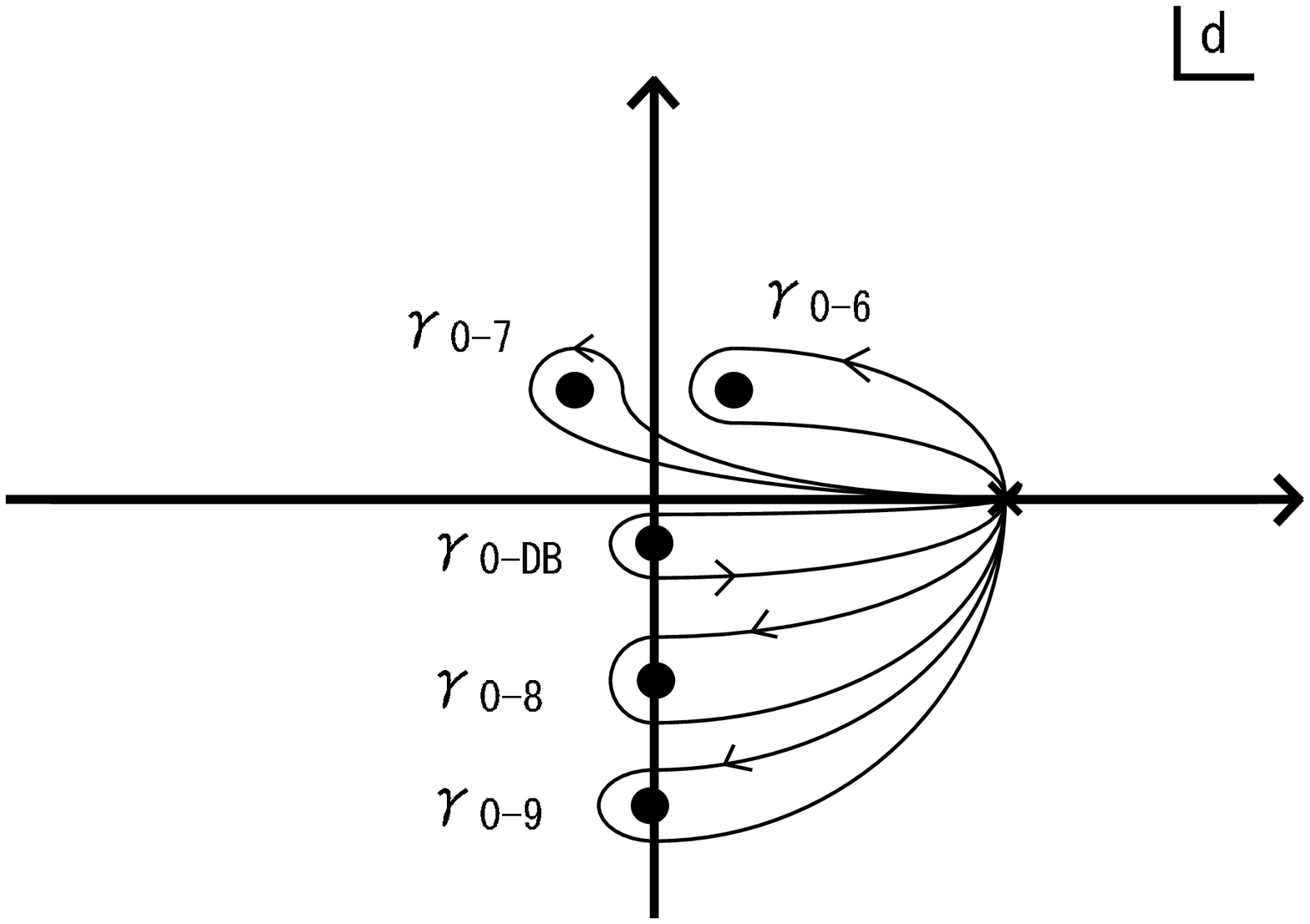} \\
 (a) & (b)
\end{tabular}
\caption{The left figure (a) shows the five branch points in the $d$-plane. The cross mark represents the base point $b_2$. The right figure (b) shows the loops corresponding to the five branch points in the $d$-plane.}
\label{fig:d}
\end{center}
\end{figure}

In the $d$-plane, 
there are four branch points $d_{0-6}, d_{0-7}, d_{0-8}, d_{0-9}$ from the first factor of \eqref{disofdis3} and a double point $d_{0-DB}$ from the second factor of \eqref{disofdis3}. They are depicted in Figure \ref{fig:d} (a). 
We label loops associated to each branch point as $\gamma_{0-6}, \gamma_{0-7}, \gamma_{0-8}, \gamma_{0-9}, \gamma_{0-DB}$ respectively as in Figure \ref{fig:d} (b). The explicit computation shows that
\bea
\rho(\gamma_{0-6})&=& \rho(\gamma_{0-7}) = \rho(\gamma_{0-8}) = \rho(\gamma_{0-9}) = W_{C_{A76}+ C_{-\theta}},\\
\rho(\gamma_{0-DB}) &=& 
\mathbb{I}, \label{trivial2}
\eea
where $W_{C_{A76} + C_{-\theta}}$ denotes the Weyl reflection with respect to the root $C_{A76} + C_{-\theta}$
\bea
\tilde{C}_{A65} &=& C_{A65} + C_{A76} + C_{-\theta},\\
\tilde{C}_{A76} &=& -C_{-\theta},\\
\tilde{C}_{-\theta} &=& -C_{A76}
\eea
Hence, the monodromy in the $d$-plane is indeed reduced to $\Z_2$. 
Note that the remaining generator is different from the one obtained from the parametrization \eqref{case1}. In this case also, the monodromy associated to a loop of the branch point from the squared factor in \eqref{disofdis3} gives the trivial monodromy. The trivial monodromy is also not a consequence of a Weyl reflection twice. In fact, the mutually local 7-branes A8 and D rotate by themselves along the loop $\gamma_{0-DB}$ and the monodromy $\rho(\gamma_{0-DB})$ becomes trivial. 



\subsection{Decomposition in $SU(5)$ gauge theory}
\label{sec:su5}

We have seen that the the monodromy group $S_3$ obtained in the $a_0$-plane gets reduced to $\Z_2$ by the parametrization \eqref{case1} and \eqref{case2-1}, \eqref{case2-2} at least in the subspace, in the $a_{-3}$-plane and the $d$-plane respectively. In order for the reduction of the monodromy in those cases, the factorization of the monodromy locus with a squared factor such as \eqref{disofdis1} and \eqref{disofdis3} is crucial. In the case of an $SU(5)$ gauge theory, we will only check the factorization of the monodromy locus. 

In order to have a $SU(5)$ gauge symmetry on 7-branes, $Y_1$ needs the $A_4$ singularity. The defining equation of the $Y_1$ has been already written in \eqref{defeq} which has the $A_4$ singularity at $(x, y) = (\frac{a_5^2}{12}, 0)$. One can easily compute the discriminant of \eqref{defeq}. The form of the discriminant is schematically
\be
\Delta = z^{5}\Delta_{SU(5)}^{\prime}(z;a_0, a_2, a_3, a_4, a_5, f_0, g_0). \label{dis5}
\ee
The $A_4$ singularity geometrically realizes five D7-branes which are described by the $z^5$ factor of \eqref{dis5}. The five D7-branes are located at $z=0$. The locations of the other seven 7-branes are described by $\Delta_{SU(5)}^{\prime}$, a degree seven equation in $z$. 
The branch points of the motion of the seven 7-branes are characterized by the discriminant $\tilde{\Delta}_{SU(5)}^{\prime}$of the degree seven equation $\Delta_{SU(5)}^{\prime}$. The schematic form of $\tilde{\Delta}_{SU(5)}^{\prime}$ is 
\be 
\tilde{\Delta}_{SU(5)}^{\prime} = 16777216 a_5^5 A^{\prime}{^3} B^{\prime}, \label{disofdis5}
\ee
where $A^{\prime}$ and $B^{\prime}$ are complicated function of $a_r (r=0, 2, 3, 4, 5), f, g$.


Let us first check whether the monodromy locus \eqref{disofdis5} becomes a factorized form or not if one applies the parametrization \eqref{a3factor} of the case I to \eqref{disofdis5}. The result for the simple ansatz is 
\be
\tilde{\Delta}^{\prime}_{SU(5)} = 16777216 a_5^5 A^{\prime}{}^3 (-2 a_{-3} a_2 + 2a_{-3}^2 a_5 + a_3 f)^2 C^{\prime}, \label{disofdis5-1}
\ee
where $C^{\prime}$ is a function of $a_r, (r=-3, 2, 3, 4, 5), f ,g$. 
Note that the factorization with squared factors
\be
B^{\prime} \rightarrow (-2 a_{-3} a_2 + 2a_{-3}^2 a_5 + a_3 f)^2 C^{\prime},
\ee
occurs in the case of the parametrization \eqref{a3factor}. This factorization behavior is very similar to the case of \eqref{disofdis1}. We may expect a reduction of the monodromy because of the squared factor in \eqref{disofdis5-1}. 


Next we apply the Higgs bundle analogy ansatze \eqref{14-0}--\eqref{14-5} and \eqref{holtuning} to \eqref{disofdis5}. The result for the Higgs bundle analogy ansatz is 
\be
\tilde{\Delta}^{\prime}_{SU(5)} = 16777216 d_4^5 b_1^5 (F A^{\prime \prime})^3 (d_4 b_0^6 - d_3 b_0^5 b_1 + d_2 b_0^4 b_1^2 + 2 d b_0^3 b_1^4 + d_4 b_0^2 b_1^6 F + d_3 b_0 b_1^7 F + d_2 b_1^8 F)^2 D^{\prime}, \label{disofdis5-2}
\ee
where $D^{\prime}$ is a function of $d, d_r, (r=2, 3, 4), b_0, b_1, F$. The factorization with squared factors
\be
B^{\prime} \rightarrow (d_4 b_0^6 - d_3 b_0^5 b_1 + d_2 b_0^4 b_1^2 + 2 d b_0^3 b_1^4 + d_4 b_0^2 b_1^6 F + d_3 b_0 b_1^7 F + d_2 b_1^8 F)^2 D^{\prime}, \label{disofdis5-2-1}
\ee 
also occurs in this case. We observe that a squared factor appears in \eqref{disofdis5-2-1} as in the case of \eqref{disofdis3}. Hence, we may also expect a reduction of the monodromy. 

We can also compute the discriminant \eqref{disofdis5} for the parametrization \eqref{23-0}--\eqref{23-5} and \eqref{su2u3hol} of the $S(U(2) \times U(3))$ spectral cover. Then, the monodromy locus \eqref{disofdis5} becomes
\be
\tilde{\Delta}^{\prime}_{SU(5)} = 16777216 (b_2 d_3)^5 A^{\prime}{}^3 (E^{\prime} F^{\prime} G^{\prime}{}^2), \label{disofdis5-3}
\ee
where $E^{\prime}, F^{\prime}, G^{\prime}$ are functions of $b_2, b, d_2, d_3, d, e_0, e_1, F$. Note that we also have a factorization $B^{\prime} \rightarrow E^{\prime} F^{\prime} G^{\prime}{}^2$ and have a factor squared. The appearance of the squared form may also indicate the reduction of the monodromy. 


The last example is the parameterization for the $S(U(1) \times U(1) \times U(3))$ spectral cover. If one inserts the parameterization \eqref{113-0}--\eqref{113-5} and the tuning for $f$ and $g$ in \eqref{holtuning} into \eqref{disofdis5}, then we have 
\be
\tilde{\Delta}^{\prime}_{SU(5)} = -16777216 (2 b_0^3 b_1^2 d_3)^5 (2 b_1^2 F A^{\prime \prime \prime})^3 (b_1^2 (d_3 b_0 + d_2 b_1)^2 (3b_0^8 - 6 b_0^4 b_1^6 F - b_1^{12} F^2)^2 H^{\prime}{}^{2} I^{\prime}),
\ee 
where $H^{\prime}$ and $I^{\prime}$ are complicated functions of $b_0, b_1, d, d_2, d_3, F$. Hence, we also obtain four squared factors from $B^{\prime}$
\be
B^{\prime} \rightarrow b_1^2 (d_3 b_0 + d_2 b_1)^2 (3b_0^8 - 6 b_0^4 b_1^6 F - b_1^{12} F^2)^2 H^{\prime}{}^{2} I^{\prime}.
\ee
Hence, it is natural to expect that the monodromy is more reduced compared with the cases of \eqref{disofdis5-1}, \eqref{disofdis5-2} and \eqref{disofdis5-3}. This is consistent with the fact that we have two unbroken $U(1)$ symmetries in heterotic string theory and it indicates we would have two, not one, monodromy invariant two-cycles.

Therefore, the appearance of the squared factor in $B^{\prime}$ in \eqref{disofdis5} is ubiquitous for all the decompositions considered in this paper, which is very non-trivial. 
Also, the tuning of $f, g$ is crucial in these cases. If we do not impose the tuning of $f, g$ and only use the tuning for the spectral data $a_r, (r=0,2,\cdots, 5)$, such factorizations would never occur. 





\section{Conclusion}

In this paper, we have discussed spectral covers for a rank five vector bundle whose structure group, a subgroup of $SU(5)$, is  decomposed such that it contains $U(n)$-type groups. This type of vector bundle is very important for constructing realistic models from heterotic string compactifications from the reasons explained in the introduction. 
We have explicitly found the parameterizations for the $S(U(1) \times U(4))$, $S(U(2) \times U(3))$ and $S(U(1) \times U(1) \times U(3))$ spectral surface equations in heterotic string compactifications. 

Although those spectral covers can be obtained by requiring certain special parameterizations for the $SU(5)$ spectral cover equation, the parameterizations we have found have important differences from the parameterizations for the decomposed spectral surface of the Higgs bundle in F-theory compactifications. Our solutions require extra tunings for the complex structure moduli of the elliptic fibered Calabi--Yau threefold $Z_3$. This is essentially because a global section of the elliptic fibration except for the zero section is necessary for supporting a point which is the group sum of the points from $U(n)$-type spectral cover. The existence of the global section can be achieved by the special complex structure moduli of the elliptic fiber. 

In F-theory compactifications, this kind of tuning maps to the tuning for the sections which is higher order in the expansion of the normal coordinates to the GUT 7-branes compared with the sections which parameterize the decomposition spectral surface of the Higgs bundle. This is consistent with the result of \cite{Hayashi:2010zp}. Ref.~\cite{Hayashi:2010zp} has shown that the factorization for the spectral cover of the Higgs bundle is not enough for ensuring an unbroken $U(1)$ symmetry in a low energy effective theory, and the tuning for the higher order terms would be also necessary. This tuning is indeed predicted from the analysis of the decomposition of the spectral cover in heterotic string compactifications. For the parameterization of the $S(U(1) \times U(4))$ spectral cover, the solution in analogy with the Higgs bundle can be used. However, the parameterization of the $S(U(2) \times U(3))$ spectral cover has different dependence on parameters for $a_0$ compared with the decomposed spectral surface of the Higgs bundle picture. 


We also discussed the presence of the unbroken $U(1)$ symmetry in F-theory compactifications by applying the parameterizations we have found in heterotic string compactifications. In F-theory compactifications, the existence of the unbroken $U(1)$ symmetry is translated into the existence of the monodromy invariant two-cycle in the K3 fiber of the Calabi--Yau fourfold $X_4$. In order to compare with the heterotic string analysis, we take the stable degeneration limit of $X_4$, and compute the monodromy invariant two-cycle in the rational elliptic surface fiber. For simplicity, we considered $E_6$ gauge theories and apply the $(1+2)$ decomposition of the case I and the case II for the $SU(3)_{\perp}$ structure group. Without any tuning, we have checked that the monodromy group in the $a_0$-plane is $S_3$, which is the Weyl group of $SU(3)_{\perp}$. On the other hand, we have found that the monodromy is reduced to $\Z_2$ for both case I and case II at least in the $a_{-3}$-plane, $d$-plane respectively by the application of the parameterizations of the case I and the case II. In the subspace, we have obtained the trivial monodromy from the squared factors of the monodromy loci. We have only computed the monodromy in a subspace, and one needs the computation in a full moduli space to ensure the true monodromy group. However, we expect that the results we have found may not change by generalizing the analysis into the full moduli space since the monodromy locus is globally factorized and contains the squared factors. Certainly, it would be interesting to extend our analysis to the monodromy of the full moduli space, and check the expectation.

One may expect that the monodromy from the squared factor becomes trivial since it would give a Weyl reflection twice. However, the trivial monodromy of \eqref{trivial1} and \eqref{trivial2} was actually obtained in a more non-trivial way. The trivial monodromy \eqref{trivial1} was obtained when the mutually non-local A7, A8, and D 7-branes rotate by more than $2\pi$ radian. The trivial monodromy \eqref{trivial2} was obtained when mutually local A8 and D 7-branes rotate together by themselves. Hence, none of the trivial monodromy is a consequence of the Weyl reflection twice. The situation is similar to the trivial monodromy from the factor $A^3=0$ in \eqref{disofdis}. In this case also, the trivial monodromy was obtained by the rotation of the mutually non-local A7, A8 7-branes \cite{unpublished}. Therefore, one might tempt to say that the monodromy from the factor whose order is more than or equal to two may become trivial although the trivial monodromy may be given in a non-trivial way. 

An interesting extension of our work is to find a general parameterization for the decomposed spectral covers in heterotic string compactifications. As for the $S(U(1) \times U(4))$ spectral cover, the parameterization of the case II does not include the generic parameterization of the case I. There might be a general parameterization which includes both case I and case II or there might be no parameterization which include both cases.

Further generalization is possible in F-theory compactifications. Our solutions would be only valid in the stable degeneration limit or at least close to it. Even if one assumes an $E_8$ singularity on the other patch of the base $\P^1$ of the elliptic K3 fiber, we still have two more 7-branes which participate in the computation of the monodromy. The factorization of the monodromy loci is in fact not occurred by the parameterizations we have found if one includes the two 7-branes. It would be interesting to find a solution which is also valid beyond the stable degeneration limit.


\vspace*{1.2cm}
\noindent
{\bf Acknowledgments}: We would like to thank Kenji Hashimoto, Seung-Joo Lee and Taizan Watari for discussions. We would like to thank the organizers of a program ``16th APCTP Winter School on Fields and Strings" in APCTP, POSTECH, where a part of the work was done. The work of K.-S.C. was supported by the Ewha Womans University Research Grant of 2012.

\appendix

\section{Group law on torus}
\label{sec:groupsum}

In this appendix we deal with a group law on a torus or elliptic curve $E$.\footnote{Here we simply denote a single elliptic curve by $E$, whereas $E_p$ is used in the main text to contrast it from elliptic fiber.}
We have a group law on $E$ with an identity element $o$ as a point, thanks to the following theorem \cite{Shafarevich}. Consider a map from a point $Q$ on $E$ to a divisor class $C_Q$ of degree zero containing $Q-o$, where the minus `$-$' means formal subtraction of divisors. Then we have one-to-one correspondence between  $Q$ and $C_Q$. This correspondence is nontrivial because $E$ is not a rational curve, and no two points on $E$ are linearly equivalent. Denoting formal sum of divisors as `$+$', the addition `$\boxplus$' of this group is defined as $Q_1 \boxplus Q_2 = Q$ if $C_{Q_1} + C_{Q_2} = C_Q$, or
\begin{equation} \label{sum}
 Q_1+Q_2 \sim (Q_1 \boxplus Q_2)+o,
\end{equation}
where it is defined by linear equivalence relation $Sim$.
The binary operator $\boxplus$ maps two points to one, so that the degrees agree on both sides.

%
\begin{figure}[t]
\begin{center}
\begin{tabular}{ccc}
  \includegraphics[scale=0.25]{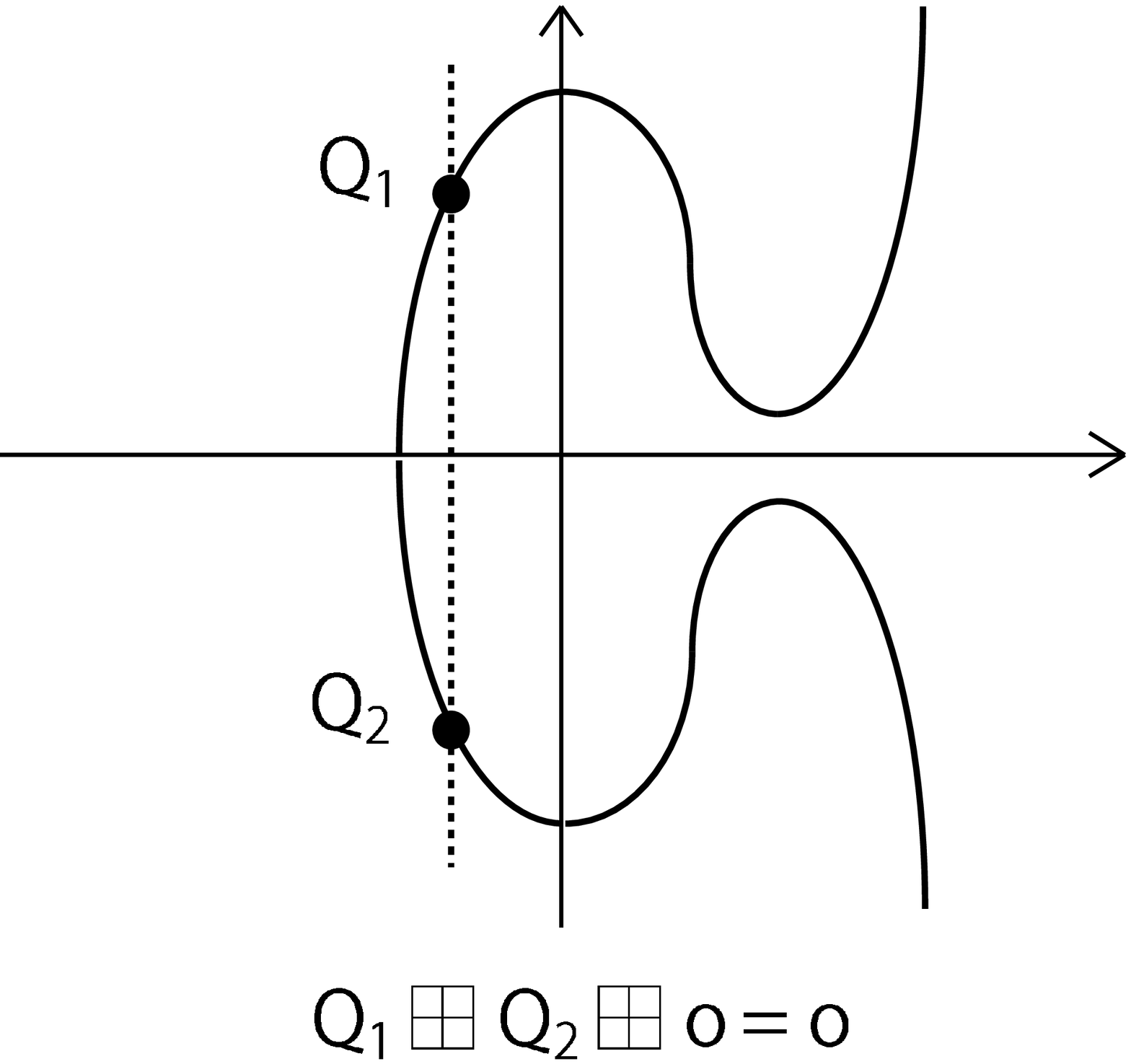} &  
  \includegraphics[scale=0.25]{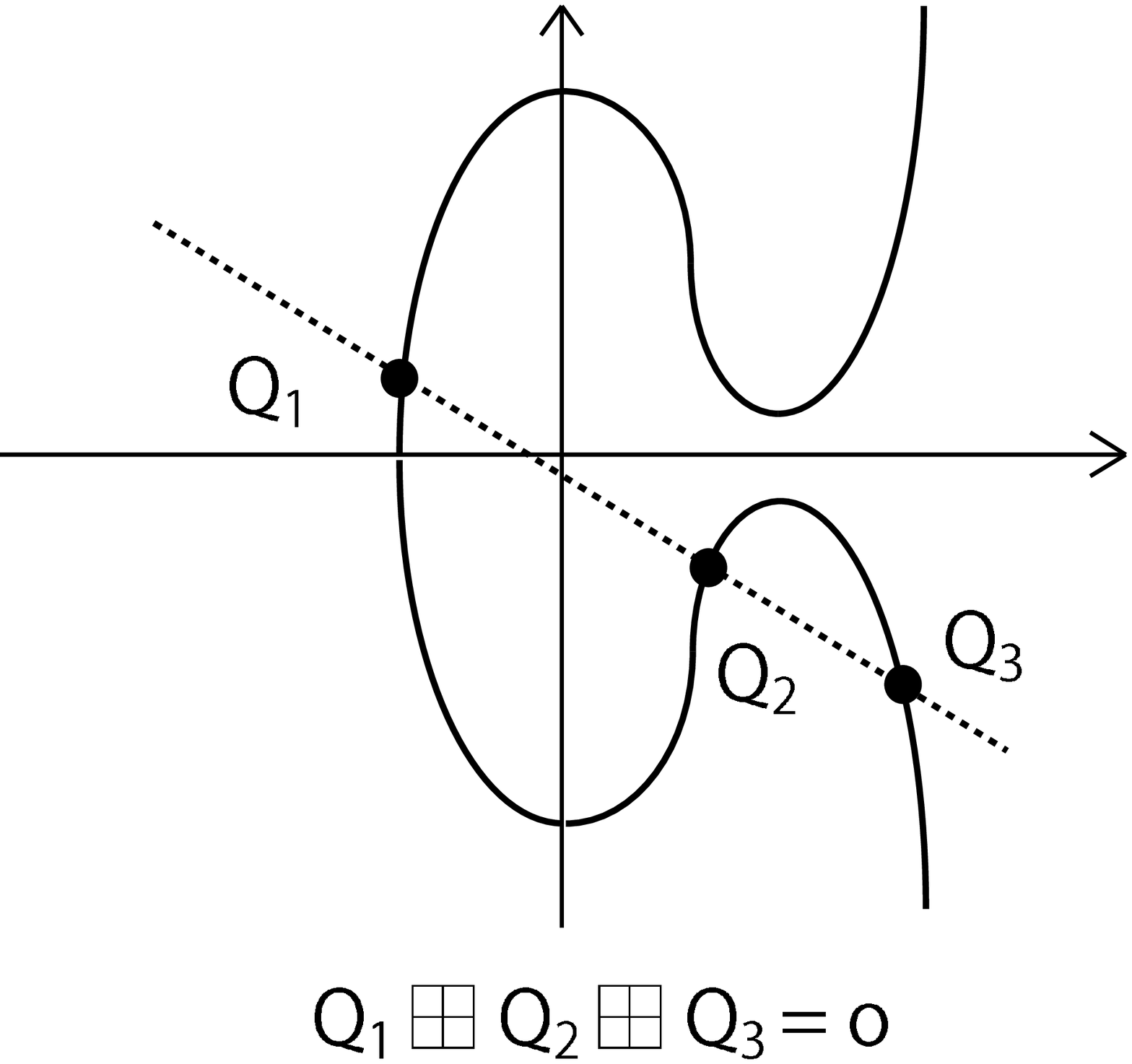} &
  \includegraphics[scale=0.25]{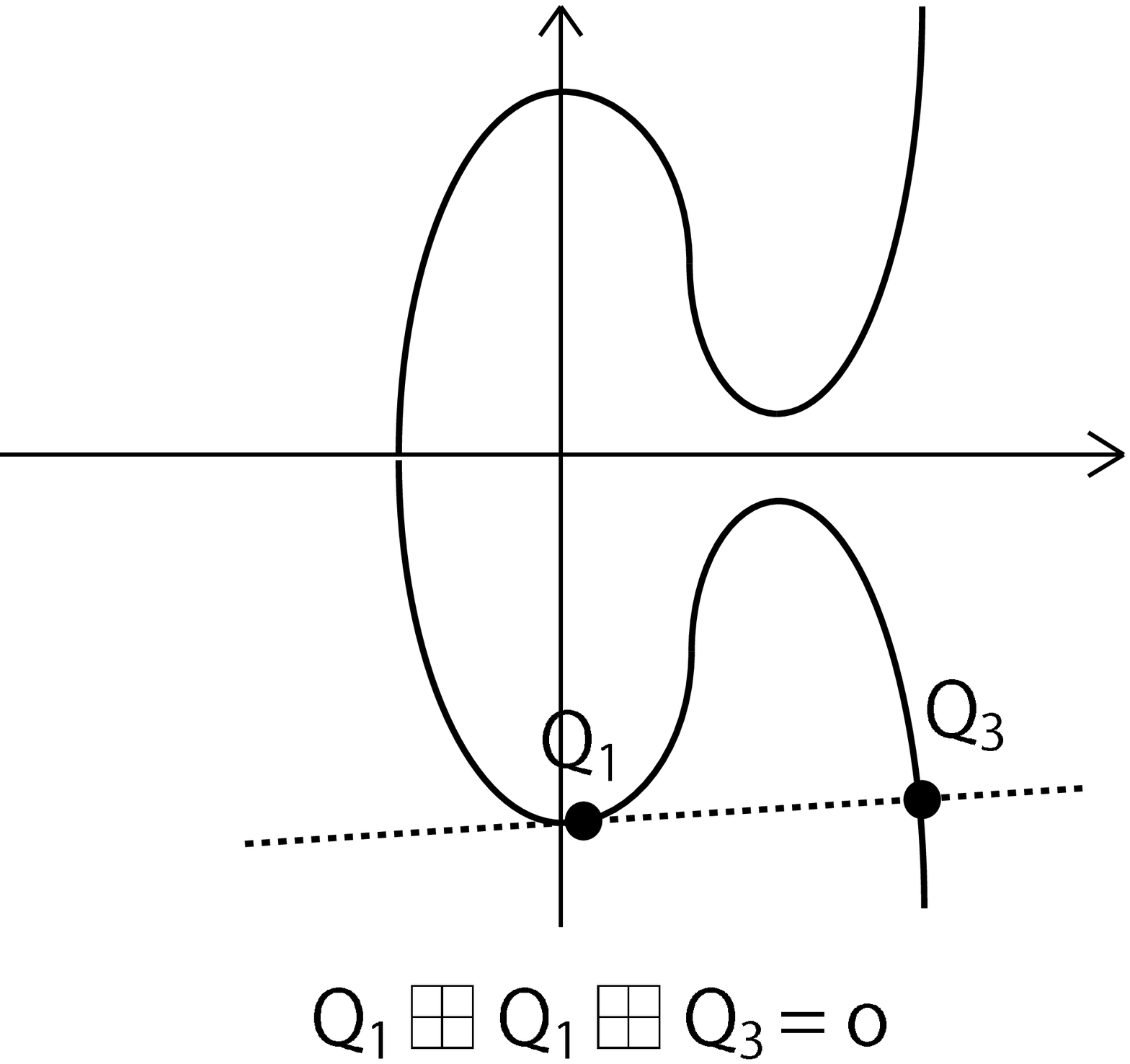}\\
 (a) & (b) & (c)
\end{tabular}
\caption{The three cases of the addition of points on an elliptic curve. The solid line denotes an elliptic curve $E$ and the dotted line denotes a line.}
\label{fig:add}
\end{center}
\end{figure}

If $E$ is described by the Weierstrass equation (\ref{Wei}), we may define a group law on $E$ very simply. This equation can be embedded in projective space $\P^2$.  With homogeneous coordinates $[X,Y,Z]$,
\begin{equation} \label{Weihom}
 ZY^2 = X^3 + f X Z^2 + g Z^3.
\end{equation}
becomes (\ref{Wei}) on the patch $Z \ne 0$, with $x=X/Z,y=Y/Z$. One point $o:[0,1,0]$ is not described on this patch, but can be regard as a `point at infinity' with $x,y$ approaching infinity. It is a special point in the sense that $o:[0,1,0]$ is always on $E$ regardless of the parameterization of $f$ and $g$. The specialty becomes more clear when one fibers the elliptic curve over a base. The origin $o$ parameterizes a global section. Also, $o$ is an inflection point, or the tangent line at $o$ has an intersection of order three. We will make use of this special property of $o$ in the following.

Now let us calculate the actual coordinates. The rules are summarized in Figure \ref{fig:add}.  By B\'ezout's theorem, a line always intersects $E$ at three generic points $Q_1, Q_2, Q_3$.
 It essentially says that, because $E$ is  described by the degree three polynomial, the common solution with the line is described by a degree three equation in $x$. $y$ is also uniquely determined by the defining equation of the line with the value of $x$. Even if these points $Q_i$ are not generic, we may apply this rule by introducing multiplicity.  So every line in $\P^2$ is linearly equivalent to a tangent line at $o$, with the multiplicity, the corresponding divisor is $3o$. The linear equivalence in $E$ can inherit the linear equivalence in $\P^2$, and we have a relation 
$$ Q_1 + Q_2 + o \sim 3 o, \ \text{ or } \  Q_2 \sim \boxminus Q_1 $$
if $Q_3 = o$ as in Figure \ref{fig:add} (a). This relation defines the minus. 
From the form of Equation (\ref{Wei}), the origin $o$ is on the lines $x=c, c\in \C$. It passes not only through $Q_1$ and $o$ but also a third point which is nothing but $Q_2 \sim \boxminus Q_1$, thus we obtain
\begin{equation}
 \boxminus (x,y) = (x,-y).
\end{equation}

By B\'ezout theorem, a line passing $Q_1$ and $Q_2$ intersects $E$ at a certain point $Q_3$, satisfying the relation 
 \begin{equation} \label{q123sum}
 Q_1 + Q_2 + Q_3 \sim 3 o, \ \text{ or }\ Q_1 \boxplus Q_2 \boxplus Q_3 \sim o.
\end{equation}
where the sum relation (\ref{sum}) is used. Associativity of group law is understood.
So we 
draw a line passing through $Q_1: (x_1,y_1)$ and $Q_2: (x_2,y_2)$ and assume $Q_1 \ne Q_2$. The third intersection between this line and $E$ gives the minus of $Q_1 \boxplus Q_2 = (x_3,-y_3)$, where
\begin{equation} \label{sumx}
 x_3 = \left( \frac{y_2-y_1}{x_2-x_1} \right)^2 -x_1 - x_2.
\end{equation}
and
\begin{equation} \label{sumy}
 y_3 = y_1 + \left(\frac{y_2-y_1}{x_2-x_1} \right) (x_3-x_1).
\end{equation}
This situation is depicted in Figure \ref{fig:add} (b).

For the case $Q_1 = Q_2$, we draw a tangent line at $Q_1$
$$ 2y_1 (y-y_1) - (x-x_1)(3x_1^2 + f) =0$$
and the third point is given by
\begin{equation} \label{tangent}
 x_2 = \frac{(3x_1^2 + f)^2}{4(x_1^3+ f x_1 +g)} -2 x_1.
\end{equation}
Obtaining $y_2$ by substituting $x_2$ back in (\ref{tangent}), we have
\begin{equation}
 Q_1 \boxplus Q_1 = (x_2,-y_2).
\end{equation}
This is expressed in Figure \ref{fig:add} (c).

\end{document}